\documentclass[twocolumn,amsmath,amssymb,aps,pra,10pt,showkeys,superscriptaddress,longbibliography
,nofootinbib
]{revtex4-1}
\usepackage{dcolumn}
\usepackage{bm}
\usepackage{mathtools, cases}
\usepackage[unicode]{hyperref}
\usepackage{graphicx}
\usepackage{times,xcolor}
\usepackage{soul}
\hypersetup{colorlinks=true,linkcolor=blue,citecolor=blue,filecolor=black,urlcolor=blue}

\renewcommand{\d}{\mathrm{d}}

\begin{document}

\title{Revised scattering exponents for a power-law distribution of surface and mass fractals}

\author{Alexander Yu. Cherny}
\email[Corresponding author, e-mail:~]{cherny@theor.jinr.ru}
\affiliation{Joint Institute for Nuclear Research, Dubna 141980, Russian Federation}

\author{Eugen M. Anitas}
\affiliation{Joint Institute for Nuclear Research, Dubna 141980, Russian Federation}
\affiliation{Horia Hulubei, National Institute of Physics and Nuclear Engineering, RO-077125 Bucharest-Magurele, Romania}

\author{Vladimir A. Osipov}
\affiliation{Joint Institute for Nuclear Research, Dubna 141980, Russian Federation}

\author{Alexander I. Kuklin}
\affiliation{Joint Institute for Nuclear Research, Dubna 141980, Russian Federation}
\affiliation{Laboratory for Advanced Studies of Membrane Proteins, Moscow Institute of Physics and
Technology, Dolgoprudniy 141700, Russian Federation}

\date{\today}

\begin{abstract}
We consider scattering exponents arising in small-angle scattering from power-law polydisperse surface and mass fractals. It is shown that a set of fractals, whose sizes are distributed according to a power-law, can change its fractal dimension when the power-law exponent is sufficiently big. As a result, the scattering exponent corresponding to this dimension appears due to the spatial correlations between positions of different fractals. For large values of the momentum transfer, the correlations do not play any role, and the resulting scattering intensity is given by a sum of intensities of all composing fractals. The restrictions imposed on the power-law exponents are found. The obtained results generalize Martin's formulas for the scattering exponents of the polydisperse fractals.
\end{abstract}


\maketitle

\section{Introduction}\label{sec:intro}

Small-angle scattering (SAS) of X-rays and neutrons is a very important tool for investigating the structural properties of partially or completely disordered systems at nano- and micro-scales~\cite{Feigin1987StructureScattering,Lindner2002NeutronsMatter}. SAS is particularly useful to describe complex hierarchical structures such as fractals \cite{Mandelbrot1982TheNature}, since it provides information that is not accessible to other methods. This information is extracted from the behaviour of the SAS intensity $I(q) \equiv (1/V^{'})\mathrm{d}\sigma/\mathrm{d}\Omega$ as a function of the scattering wave vector $q \equiv \left( 4\pi/\lambda \right) \sin \theta$. Here $V^{'}$ is the unit volume of the sample measured, $\mathrm{d}\sigma /\mathrm{d}\Omega$ is the elastic cross section, $\lambda$ is the wave length of incident radiation, and $\theta$ is half of the scattering angle.

For fractals, there is always a $q$-range over which the intensity can be described as~\cite{Feigin1987StructureScattering}:
\begin{equation}
 I(q)\propto q^{-\alpha},
 \label{eq:SASpower-law}
 \end{equation}
where $\alpha$ is called the scattering exponent. The general consensus is that when $\alpha$ is an integer, the scattering arises from a regular $d$-dimensional Euclidean object ($\alpha = 1$, $2$, and $4$ in 1d, 2d, and 3d, respectively)~\cite{Schmidt:a21839}. Otherwise the scattering is considered to arise from a fractal structure, and in this case $\alpha$ is related to the fractal (Hausdorff) dimension $D$ \cite{Mandelbrot1982TheNature,Gouyet1996PhysicsStructures,cherny11}. Simply speaking, the dimension is given by the exponent in the relation $N \propto \left( L / a \right)^{D}$ when $a \rightarrow 0$, where $N$ is the minimum number of open sets of diameter $a$ required to cover an arbitrarily set of diameter $L$.

The SAS method enables us to distinguish between mass~\cite{Teixeira1988Small-angleSystems} and surface~\cite{Bale1984Small-AngleProperties} fractals. Let us consider a two-phase geometric configuration in $d$-dimensional Euclidean space consisting of a set of dimension $D_{\mathrm{m}}$ (i.e. the phase labelled as ``mass") and of its complement of dimension $D_{\mathrm{p}}$ (i.e. the phase labelled as ``pores").  The dimension of the boundary between the two phases is denoted by $D_{\mathrm{s}}$. Then the scattering exponent is $\alpha = D_{\mathrm{s}} = D_{\mathrm{m}} < d$ with $D_{\mathrm{p}}=d$ for mass fractals, and $\alpha = 2d - D_{\mathrm{s}}$ with $d-1 \leqslant D_{\mathrm{s}} < d$ and $D_{\mathrm{m}}=D_{\mathrm{p}}=d$ for surface fractals~\cite{Martin1987ScatteringFractals,Schmidtrev91,pfeifer02}. For instance,  in two dimensions ($d=2$),
the sample is a mass fractal when $\alpha < 2$, and it is a surface fractal when $2 < \alpha < 3$. Physically, for a mass fractal, the smaller the value of $\alpha$, the lower its fractal dimension and the more open it is, while for a surface fractal, the situation is reversed. For instance, for a perfectly smooth line in $2d$, $D_{\mathrm{s}}\rightarrow 1$ and $\alpha \rightarrow 3$. When the line is so ``wriggled" that it almost fulfils the plane, $D_{\mathrm{s}} \rightarrow 2$ and $\alpha \rightarrow 2$.

However, for a system of \emph{power-law} polydisperse fractals, the scattering exponent $\alpha$ is changed for high polydispersity. Martin showed \cite{martin86} that the scattering exponent of a power-law polydisperse mass fractals lies in the interval $[0,d]$ and always depends on the mass fractal dimension. For surface fractals, the scattering exponent lies within $[0,d+1]$ and it is independent of the surface fractal dimension for a large range of the polydispersity exponent. The total scattering intensity was obtained by Martin merely as a sum of intensities, which assumes the absence of spatial correlations between the positions of separate fractals.

By extending Martin's approach, we demonstrate that for a set of fractals, whose sizes are distributed according to a power-law, the spatial correlations can play an important role. As is shown below, the scattering exponent $\alpha$ changes due to the correlations between positions of different fractals, provided the power-law exponent is sufficiently big. However, for large values of the momentum transfer, the spatial correlations do not play any role, and the resulting scattering intensity is indeed given by a sum of intensities of the composing fractals. Thus Martin's results are recovered in the range of high momentum transfer. In order to simplify numerical simulations, only two-dimensional models are considered.

The paper is organized as follows. We start in Sec.~\ref{sec:entireDim} with a mathematical background for describing the connection between the fractal dimension of a polydisperse fractal system and the exponent of power-law distribution of fractal sizes. In Sec.~\ref{sec:scat_exp} we obtain the scattering exponents for polydisperse mass and surface fractals in terms of the power-law exponent. This is followed in Secs.~\ref{sec:models}, \ref{sec:cont1}, \ref{sec:cont2}, \ref{sec:rpl-disks} and~\ref{sec:rpl-CMF} by presenting models and numerical simulations for both discrete and continuous power-law distributions of fractal sizes. Finally, in Sec.~\ref{sec:concl} we discuss the differences between our and Martin's results.

\section{The fractal dimension of the polydisperse fractal system}
\label{sec:entireDim}

In order to study the influence of the power-law polydispersity on the scattering exponent of SAS intensity, we consider a polydisperse system of non-overlapping fractals embedded in a finite region of $d$-dimensional space. The total number of fractals are supposed to be infinitely large, and their shape and structure are assumed to be the same, so each of them can be obtained from another one by uniformly scaling. Their sizes vary from 0 to $R$ and obey the power-law distribution $1/r^{\gamma+1}$,  where the power-law exponent\footnote{\label{foot1} Compared to the review \cite{Schmidtrev91}, the exponent $\gamma$ is shifted by one, which is technically more convenient.} lies in the range $0<\gamma\leqslant d$. This means that the number of fractals $\mathrm{d}N(r)$ whose sizes fall within the range ($r,r + \d r$) is proportional to $\mathrm{d}r/r^{\gamma+1}$.

The question arises, what is the fractal (Hausdorff) dimension of the entire fractal system if the dimension of each composing fractal is $D$? In order to answer this question, we introduce a lower cutoff length $a$ and consider a finite number of fractals with sizes $a\leqslant r \leqslant R$. According to the definition of fractal dimension, the minimal number of balls of radius $a$ needed to cover a fractal of size $r$ is proportional to $(r/a)^{D}$. The minimal number of balls for covering the entire system with a finite cutoff length $a$ is given by the integral
\begin{align}
N(a)\propto &
\int_{a}^{R}\mathrm{d} r\, r^{-\gamma-1} \left(\frac{r}{a}\right)^{D}= \nonumber\\
&\frac{1}{a^D}\frac{1}{D-\gamma}\left(R^{D-\gamma}-a^{D-\gamma}\right).
\label{eq:numberofballstocover}
\end{align}
When $a\to 0$, the main asymptotics of $N(a)$ depends on the sign of $D-\gamma$. If it is positive then the first term in the parentheses dominates and, hence, $N(a)\propto 1/a^{D}$. If the sign is negative, the second term dominates, and the asymptotics is given by $N(a)\propto 1/a^{\gamma}$. In accordance with the definition of Hausdorff dimension, $N(a)\propto 1/a^{D_\mathrm{tot}}$, and we obtain the fractal dimension of the entire system
\begin{align}
D_{\mathrm{tot}}=
\begin{cases}
D, &\text{for}\ \gamma\leqslant D,\\
\gamma, &\text{for} \ D\leqslant\gamma\leqslant d.
\end{cases}
\label{eq:Deff}
\end{align}
Here the inequality $\gamma\leqslant d$ is needed to avoid overlapping between the fractals. This is because the fractal dimension of the entire system cannot exceed the maximum Hausdorff dimension $d$, which corresponds to complete filling of a finite region of $d$-dimensional space.

Some caveats should be made. The above considerations suppose implicitly that the entire set of the power-law polydisperse fractals of dimension $D$ forms a \emph{fractal} of dimension $D_{\mathrm{tot}}$. Strictly speaking, this is not the case in general. This is because self-similarity in fractals implies that a vicinity of any point that belongs to a fractal is its small copy. However, a small vicinity of an inner point of each separate fractal is self-similar to this fractal but not the entire set. Nevertheless, the fractal dimension of the entire set of the polydisperse fractals is correctly given by Eq.~(\ref{eq:Deff}). Below we show that this overall set exhibits the fractal properties related to small-angle scattering.

\section{The scattering exponent of the polydisperse mass and surface fractals}
\label{sec:scat_exp}

As discussed in Sec.~\ref{sec:intro}, the scattering exponent of the small-angle scattering ($I\propto q^{-\alpha}$) is directly related to the fractal dimension \cite{Bale1984Small-AngleProperties,Sinha1984}
\begin{equation}
\alpha = \begin{cases}
   D_{\mathrm{m}}, &\text{for mass fractals},\\
   2d-D_{\mathrm{s}}, &\text{for surface fractals}.
   \end{cases}
  \label{eq:tau}
\end{equation}
Besides, the fractal dimensions always obey the inequalities $0 <D_\mathrm{m} \leqslant d$ and $d-1 \leqslant D_\mathrm{s}< d$.

\begin{figure}[!tb]
\centerline{\includegraphics[width=.85\columnwidth]{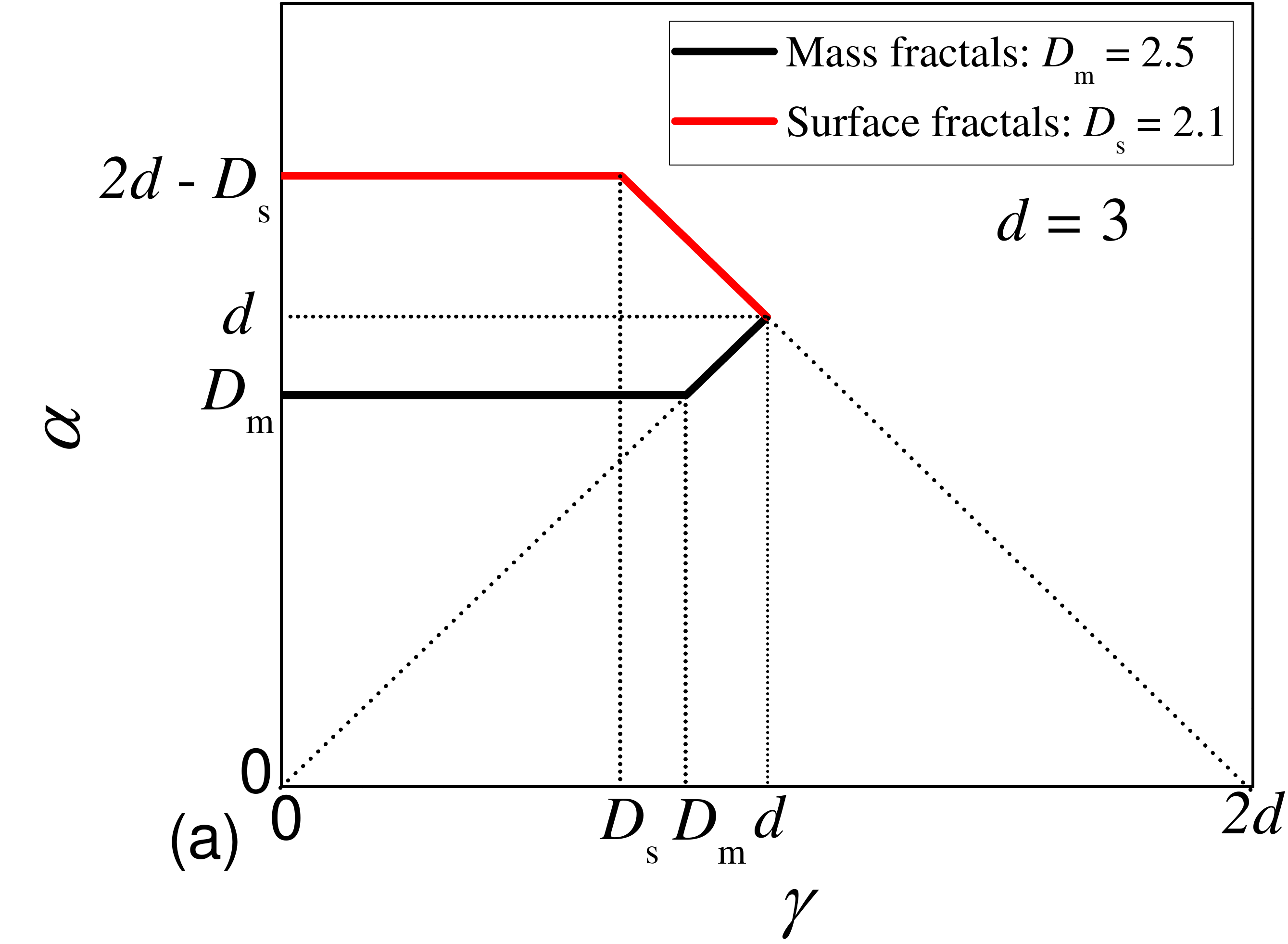}}
\centerline{\includegraphics[width=.85\columnwidth]{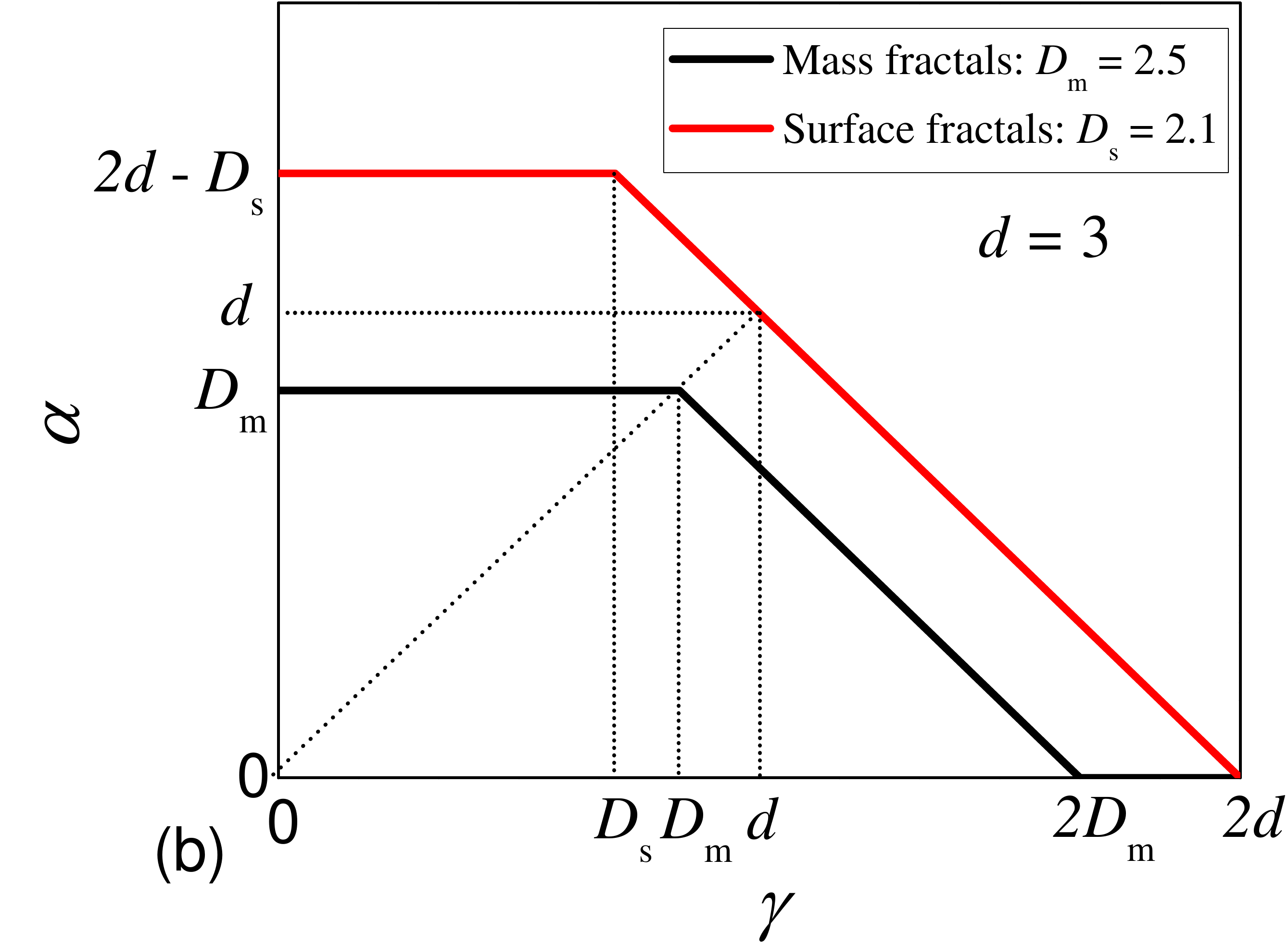}}
\caption{\label{fig:scexp}The dependence of the scattering exponent $\alpha$ on the power-law exponent $\gamma$ for 3d mass (black lower lines) and surface (red upper lines) fractals. (a) Results obtained in this paper given by Eqs.~(\ref{eq:gamM}) and (\ref{eq:gamS}). (b) Martin's results \cite{martin86} given  by Eqs.~(\ref{eq:gamMmar}) and (\ref{eq:gamSmar}).
 }
 \end{figure}

We find from Eqs.~(\ref{eq:Deff}) and (\ref{eq:tau}) the scattering exponents for the polydisperse mass and surface fractals, respectively:
\begin{align}
\alpha_{\mathrm{m}} =&
\begin{cases}
D_{\mathrm{m}}, &\text{for}\ \gamma\leqslant D_{\mathrm{m}},\\
\gamma, &\text{for} \ D_{\mathrm{m}}\leqslant\gamma\leqslant d,
\end{cases}
\label{eq:gamM}\\
\alpha_{\mathrm{s}} =&
\begin{cases}
2d-D_{\mathrm{s}}, &\text{for}\ \gamma\leqslant D_{\mathrm{s}},\\
2d-\gamma, &\text{for} \ D_{\mathrm{s}}\leqslant\gamma\leqslant d.
\end{cases}
\label{eq:gamS}
\end{align}

Let us compare these results with Martin's \cite{martin86}. He originally related the scattering exponent with the exponent of \emph{mass} distribution, but his equations can be written in terms of the exponent of \emph{size} distribution
\cite{Schmidtrev91} (see the footnote \ref{foot1}):
\begin{align}
\alpha_{\mathrm{m}} =&
\begin{cases}
D_{\mathrm{m}}, &\text{for}\ \gamma\leqslant D_{\mathrm{m}},\\
2D_{\mathrm{m}}-\gamma, &\text{for} \ D_{\mathrm{m}}\leqslant\gamma\leqslant 2D_{\mathrm{m}},
\end{cases}
\label{eq:gamMmar}\\
\alpha_{\mathrm{s}} =&
\begin{cases}
2d-D_{\mathrm{s}}, &\text{for}\ \gamma\leqslant D_{\mathrm{s}},\\
2d-\gamma, &\text{for} \ D_{\mathrm{s}}\leqslant\gamma\leqslant 2d.
\end{cases}
\label{eq:gamSmar}
\end{align}

One can see two main differences between the exponents given by Eqs.~(\ref{eq:gamM}) and (\ref{eq:gamS}) and those of Eqs.~(\ref{eq:gamMmar}) and (\ref{eq:gamSmar}), respectively, see Fig.~\ref{fig:scexp}. First, we have the restriction related to dense packing of subsets in $d$-dimensional space, discussed above in Sec.~\ref{sec:entireDim}: the resulting dimension cannot exceed $d$. This leads to the constraints $\alpha\leqslant d$ for both mass and surface fractals. Second, the scattering exponent for mass fractals should \emph{increase} as a function of the power-law exponent when $D_{\mathrm{m}}\leqslant\gamma$ until its value reaches the maximum value $D_\mathrm{m}=d$. The reason for this is that an increment of $\gamma$ increases the density of the fractal set and, as a consequence, the resulting fractal dimension.

Note that the exponent $\alpha$ cannot be negative even when $2d-\gamma<0$, but it remains zero in this case. We study this regime numerically in Secs.~\ref{sec:cont1} and \ref{sec:cont2} below.

\section{Discrete power-law distribution: Cantor surface fractals consisting of Cantor mass fractals}
\label{sec:models}

\subsection{Model}
\label{sec:modeldiscrete}

As was shown by the authors \cite{Cherny2017ScatteringFractals,Cherny2017Small-angleSnowflake,cherny19rev}, the ``discrete'' power-law distribution can be realized as a superposition of various iterations of the generalized Cantor mass fractals (CMF) \cite{Cherny2010ScatteringFractals}. This model is easier to handle numerically than the continuous power-law distribution, since it enables an analytical representation of the scattering amplitude.

By analogy with the previous papers \cite{Cherny2010ScatteringFractals, cherny11,Cherny2017ScatteringFractals,Cherny2017Small-angleSnowflake,cherny19rev}, we consider a set of clusters, which are completely uncorrelated in positions and orientations.  For instance, this is realized when the set is sufficiently diluted in space. Below a single cluster is described.

We consider a construction in a plane ($d=2$), which realizes the discrete analog of the power-law distribution of ``building blocks'' with the exponent $\gamma$ (see Sec. 4 of Ref.~\cite{Cherny2017ScatteringFractals}). We put into the plane one block of size $h$, $k$ similar blocks of size $\beta h$, $k^2$ blocks of size $\beta^2 h$, and so on. The scaling factor is given by $\beta=k^{-1/\gamma}$. Here and below we specify the number of blocks generating at each iterations: $k=4$. In paper \cite{Cherny2017ScatteringFractals}, the composing blocks were disks.

\begin{figure}[!tb]
\centerline{\includegraphics[width=.9\columnwidth,trim={1cm 3cm 0.1cm 0.1cm}]{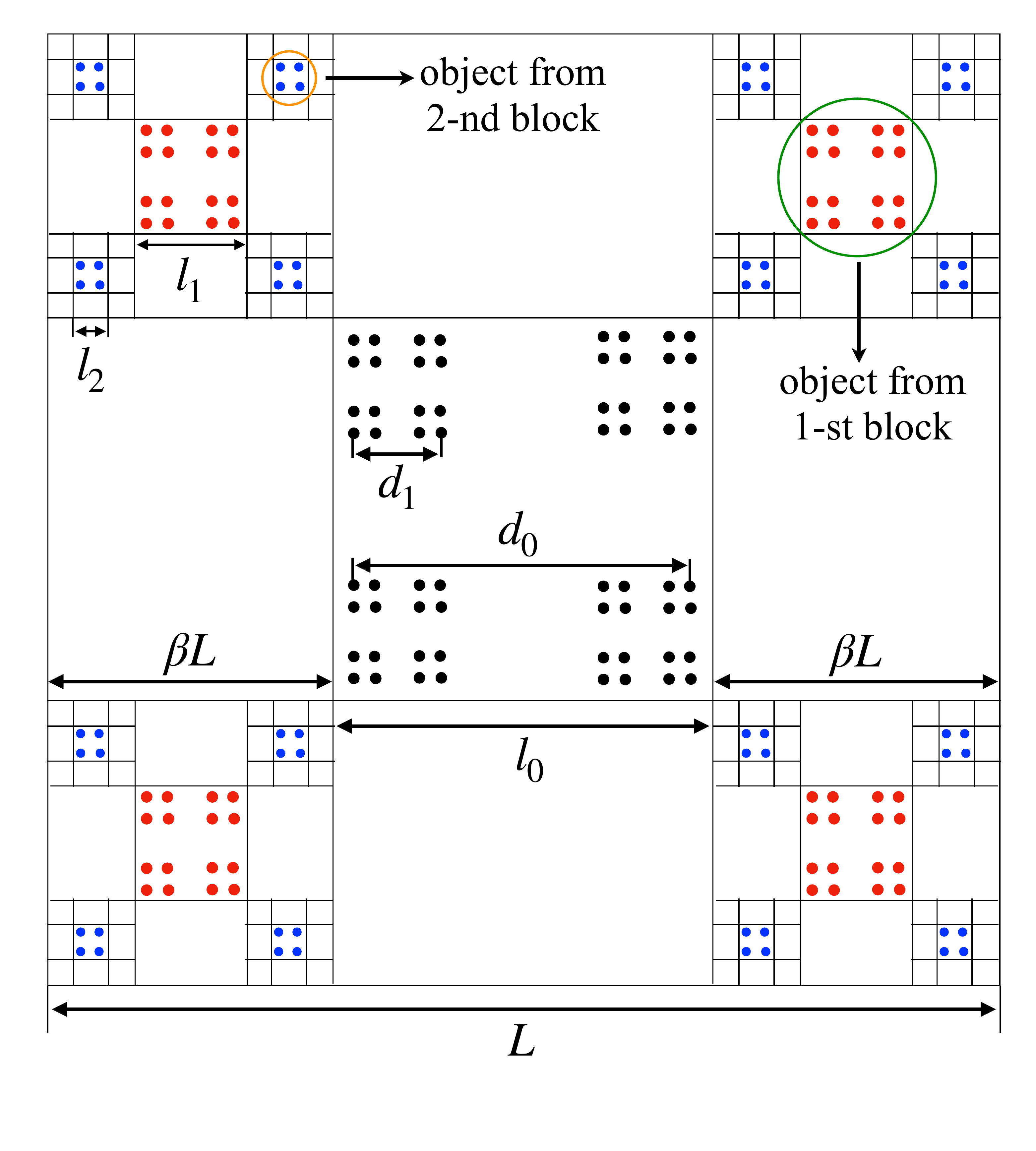}}
\caption{\label{fig:discretemodel} Construction of a system with discrete power-law distribution: CMF of scaling factor $\beta_{\mathrm{m}} = 0.33$, fractal dimension $D_{\mathrm{m}} = 1.25$, and the smallest iteration $m = 1$ is distributed with the exponent $\gamma=1.15$ corresponding to the scaling factor $\beta = 0.3$. The number of the ``block'' iterations is given by $n = 2$, and, besides, $f=1$. Black dots: scattering units from $0$th block (see the main text for details). Red dots: scattering units from $1$st block. An object from this block is encircled in the green circle. Blue dots: scattering units from $2$nd block. An object from this block is encircled in the orange circle. $d_{0}=l_{2}/\beta_{\mathrm{m}}^{3}$ and $d_{1}=l_{2}/\beta_{\mathrm{m}}^{2}$ and are the overall lengths of CMF at iterations three and two, respectively. $L$ is the overall size of the construction, and $l_{i}$, with $i = 0,1,2$ are the sizes of objects from $i$th block (see the main text).
}
 \end{figure}

Here we consider a similar construction where the primary composing blocks are replaced by mass fractals (see Fig.~\ref{fig:discretemodel}). These fractals can be the generalized Cantor fractals with a different scaling factor, which we denoted by $\beta_{\mathrm{m}}$. The scaling factor is related to the mass fractal dimension by the equation $\beta_{\mathrm{m}}=k^{-1/D_{\mathrm{m}}}$. The restrictions $\gamma< d$ and $D_{\mathrm{m}}<d$ lead to $\beta<1/2$ and $\beta_{\mathrm{m}}<1/2$, respectively.

The construction (Fig.~\ref{fig:discretemodel}) is composed from $n+1$ blocks, and the $i$th  block \st{is} consists of $k^i$ objects with the same size $l_{i}=Lf(1-2\beta)\beta^i$, $i=0,1,\cdots,n$. Here $L$ is the overall size of the construction, and $f$ is an additional dimensionless factor to control the sizes. It obeys the restriction $0<f\leqslant 1$. The $i$th object is the mass fractal of iteration $j$ depending on $i$. The $j$th mass fractal iteration consists of $k^j$  points,  where each point is supposed to have unit mass and unit scattering amplitude. We assume that the smallest $n$th object amounts to the $m$th  mass fractal iteration of the size $l_{n}=Lf(1-2\beta)\beta^n$. Then the object corresponding to the $i$th block contains the $j$th mass fractal iteration of size $l_{n}/\beta_{\mathrm{m}}^p$, where  $j=m+p(i)$ and
\begin{align}\label{pi}
p(i)=\left\lfloor(n-i)\frac{D_{\mathrm{m}}}{\gamma}\right\rfloor.
\end{align}
Here the symbol $\lfloor x \rfloor$ stands for the floor function, that is, the greatest integer less than or equal to $x$.

Such a complicated construction is needed since we build the mass fractal ``from bottom up'' with the fixed size of the smallest unit and, therefore, the mass fractal sizes can be changed only in discrete steps. If the size of the $m$th iteration is $l_{n}=Lf(1-2\beta)\beta^n$ then the higher iterations are of sizes $l_{n}/\beta_{\mathrm{m}}^p$ with $p=1,2,\cdots$. The number $p$ for $i$th block take its possible maximum value in order to ``inscribe'' the biggest mass fractal into the square bounding the $i$th block.

\subsection{Small-angle scattering properties}

With the methods developed in the papers \cite{Cherny2017ScatteringFractals,cherny19rev}, one can write down the normalized scattering amplitude of the entire construction as
\begin{align}
  F(\bm{q})=  \frac{\sum_{i=0}^{n}k^{i+p(i)}G_{i}(\bm{q})F^{\text{C}}_{m+p(i)}(\beta^{i}\bm{q})}{\sum_{i=0}^{n}k^{i+p(i)}},\label{Ftotal}
\end{align}
where $G_i(\bm{q}) =G_1(\bm{q}) G_1(\beta\bm{q}) \cdots G_1(\beta^{i-1}\bm{q})$ for $i=1, \cdots,n$ and $G_0(\bm{q})=1$, and $F^{\text{C}}_{j}(\bm{q})$ is the amplitude of the $j$th iteration of the CMF: $F^{\text{C}}_{0}(\bm{q}) =1$ and $F^{\text{C}}_{j}(\bm{q}) =G_1(\bm{q}) G_1(\beta_{\mathrm{m}}\bm{q}) \cdots G_1(\beta_{\mathrm{m}}^{j-1}\bm{q})$ for $j\geqslant1$. The generating function $G_1$ is given by $G_1(\bm{q})=\cos\frac{q_x L(1-\beta)}{2}\cos\frac{q_y L(1-\beta)}{2}$. By replacing the CMF amplitudes by one, we arrive at the structure-factor amplitude of the entire construction
\begin{align}
  G(\bm{q})=  \frac{\sum_{i=0}^{n}k^{i+p(i)}G_{i}(\bm{q})}{\sum_{i=0}^{n}k^{i+p(i)}}.\label{FStotal}
\end{align}

Then the total intensity and the structure factor can be written down through the averages over all directions of the momentum transfer $\bm{q}$:
\begin{align}
  I(q) =\langle |F(\bm{q})|^2\rangle, \quad   S(q) =&\langle |G(\bm{q})|^2 \rangle, \label{I_S}
\end{align}
respectively. The long-range asymptotics of the structure factor can be obtained from Eqs.~(\ref{FStotal}) and (\ref{I_S}) if we take into account the asymptotics \cite{cherny11} $\langle G_{i}(\bm{q})G_{j}(\bm{q})\rangle \simeq k^{-i}\delta_{ij}$ with $\delta_{ij}$ being the Kronecker symbol:
\begin{align}
S_{\mathrm{as}}=\frac{\sum_{i=0}^{n}k^{i+2p(i)}}{\left(\sum_{i=0}^{n}k^{i+p(i)}\right)^2}.\label{Ias}
\end{align}

The amplitude (\ref{FStotal}) corresponds to a set of point-like objects with different weights. The $i$th term in the sum is the normalized structure-factor amplitude of the $i$th Cantor mass fractal weighed by the factor $k^{i+p(i)}$. This construction is nothing else but the fractal-like structure (see the discussion in Sec.~\ref{sec:entireDim}) with the dimension $D_{\mathrm{tot}}$, and, hence, its scattering intensity can be described qualitatively as follows \cite{cherny11}. It is equal to one in the Guinier range  $q\lesssim 2\pi/L$, and then it falls off with the exponent $-D_{\mathrm{tot}}$ within the range $2\pi/L\lesssim q \lesssim q_{\mathrm{as}}$, and finally takes the constant value $S_{\mathrm{as}}$ when $q \gtrsim q_{\mathrm{as}}$. Once its asymptotics (\ref{Ias}) is known, the upper border of the fractal range can easily be estimated as
\begin{align}\label{eq:qas}
q_{\mathrm{as}}\simeq S_{\mathrm{as}}^{-1/D_{\mathrm{tot}}}/L.
\end{align}

For deterministic fractal structures the corresponding scattering intensity is characterized by a generalized power-law decay, i.e. a succession of maxima and minima superimposed on a simple power-law decay \cite{Cherny2010ScatteringFractals,cherny11}. In order to properly estimate the scattering exponent $\alpha$, we need to smooth the intensity without changing the value of the exponent. The smoothing can be attributed to an additional polydispersity, which does not change the exponent, or to the resolution of measuring device.
Without loss of generality, we use a log-normal distribution, defined as
\begin{equation}
D_{\mathrm{N}}(L) = \frac{1}{\sigma L (2\pi)^{1/2}}\exp\left( -\frac{[\ln(L/\mu_{0})+\sigma^{2}/2]^{2}}{2\sigma^{2}} \right),
    \label{eq:DN}
\end{equation}
where $\sigma = [\ln(1+\sigma_{\mathrm{r}}^{2})]^{1/2}$. The quantities $\mu_{0}$ and $\sigma_{\mathrm{r}}$ are the mean length and relative variance, i.e. $\mu_{0} \equiv \langle L \rangle_{D}$ and $\sigma_{\mathrm{r}} \equiv \left(\langle L^{2} \rangle_{D} - \mu_{0}^{2} \right)^{1/2}/\mu_{0}$, and $\langle \cdots \rangle \equiv \int_{0}^{\infty} \cdots D_{\mathrm{N}}(L)\mathrm{d}L.$
Then the smoothed scattering intensity  is given by \cite{Cherny2010ScatteringFractals,cherny11}
\begin{equation}
I_\mathrm{sm}(q) = \int_{0}^{\infty} I(q) D_{\mathrm{N}}(L)\mathrm{d}L.
    \label{eq:polyI}
\end{equation}

\begin{figure}[!t]
\centerline{\includegraphics[width=\columnwidth,trim={0 1cm 2.5cm 2cm}]{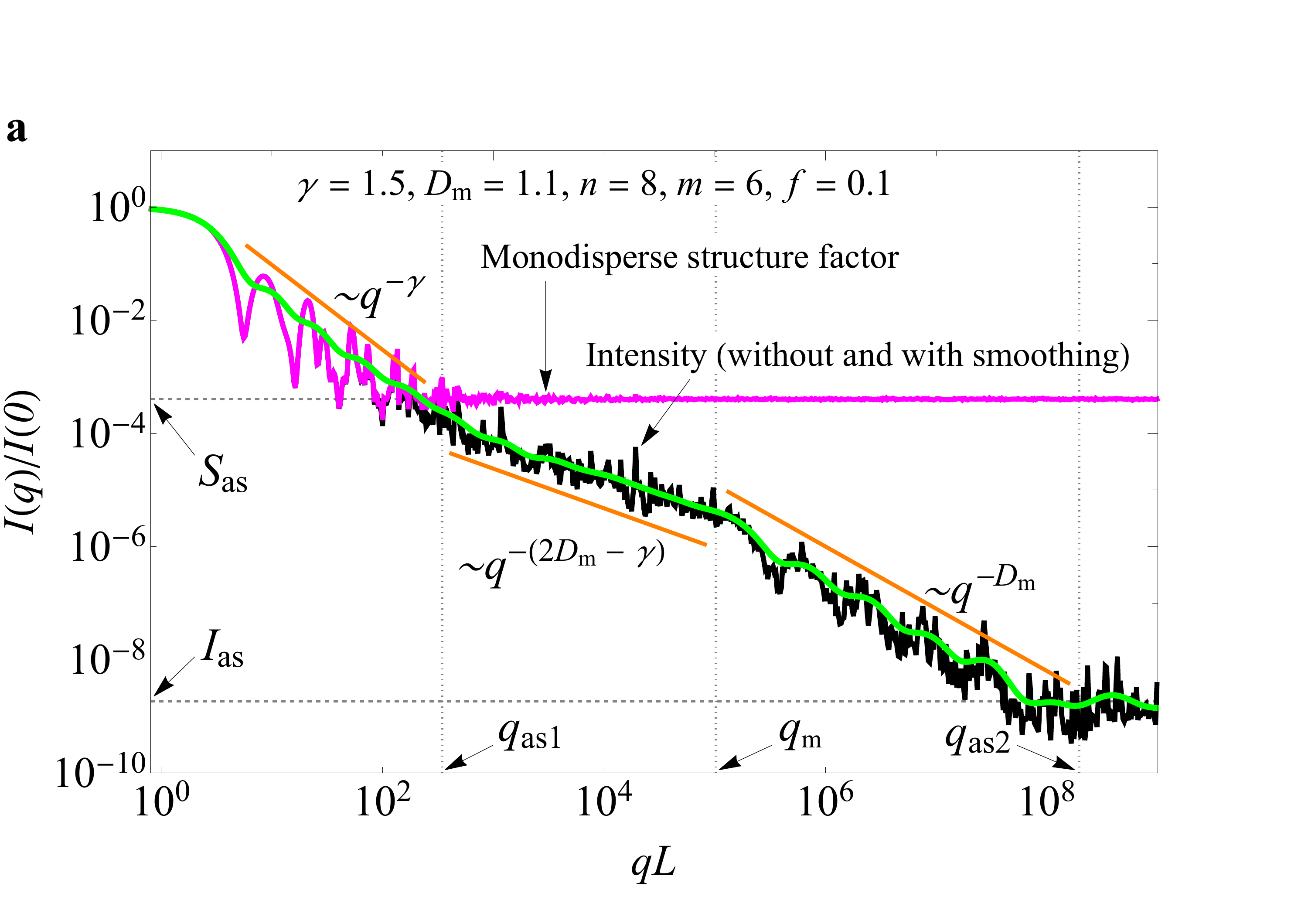}}
\centerline{\includegraphics[width=\columnwidth,trim={0 1cm 2.5cm 2cm}]{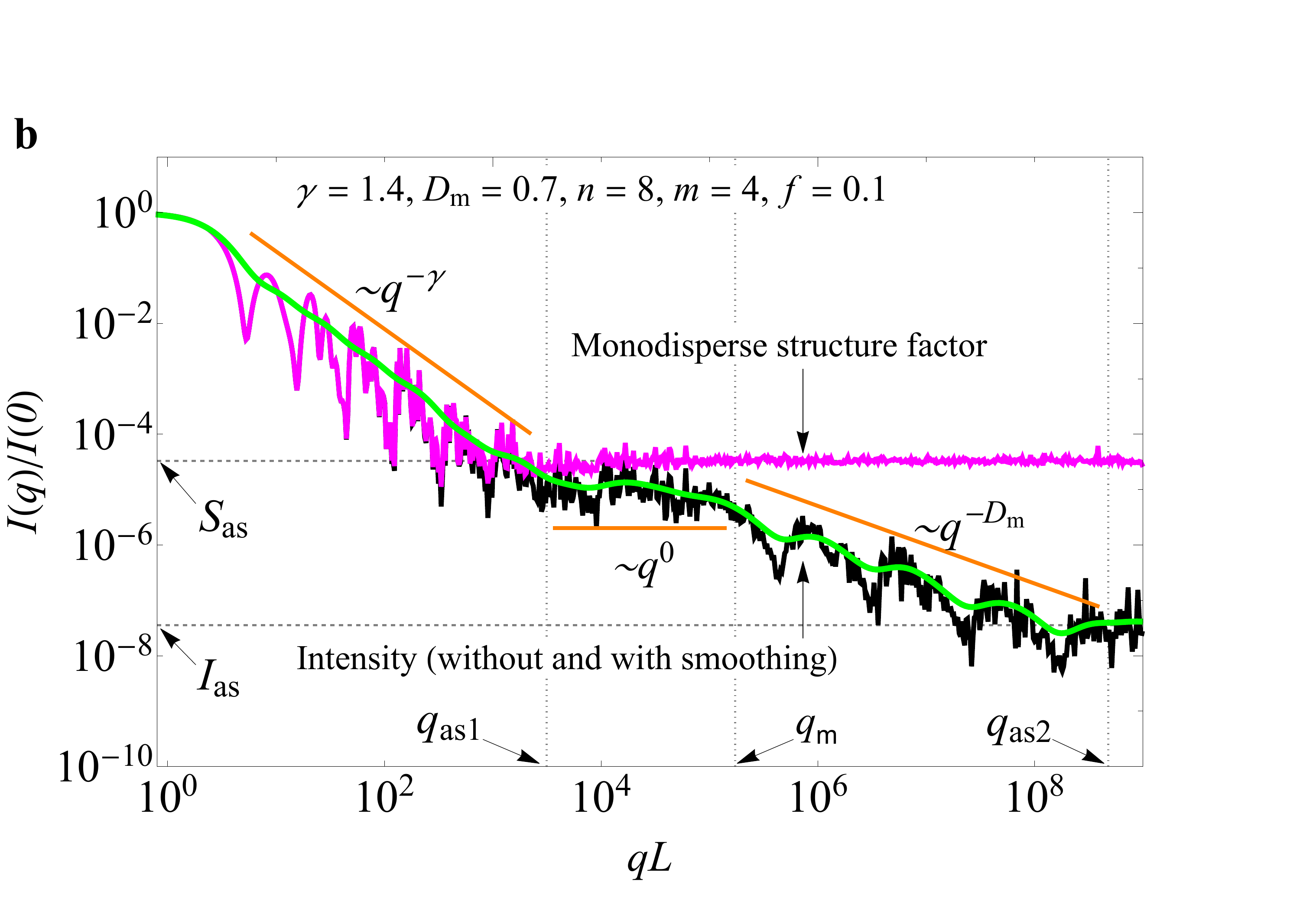}}
\centerline{\includegraphics[width=\columnwidth,trim={0 1cm 2.5cm 2cm}]{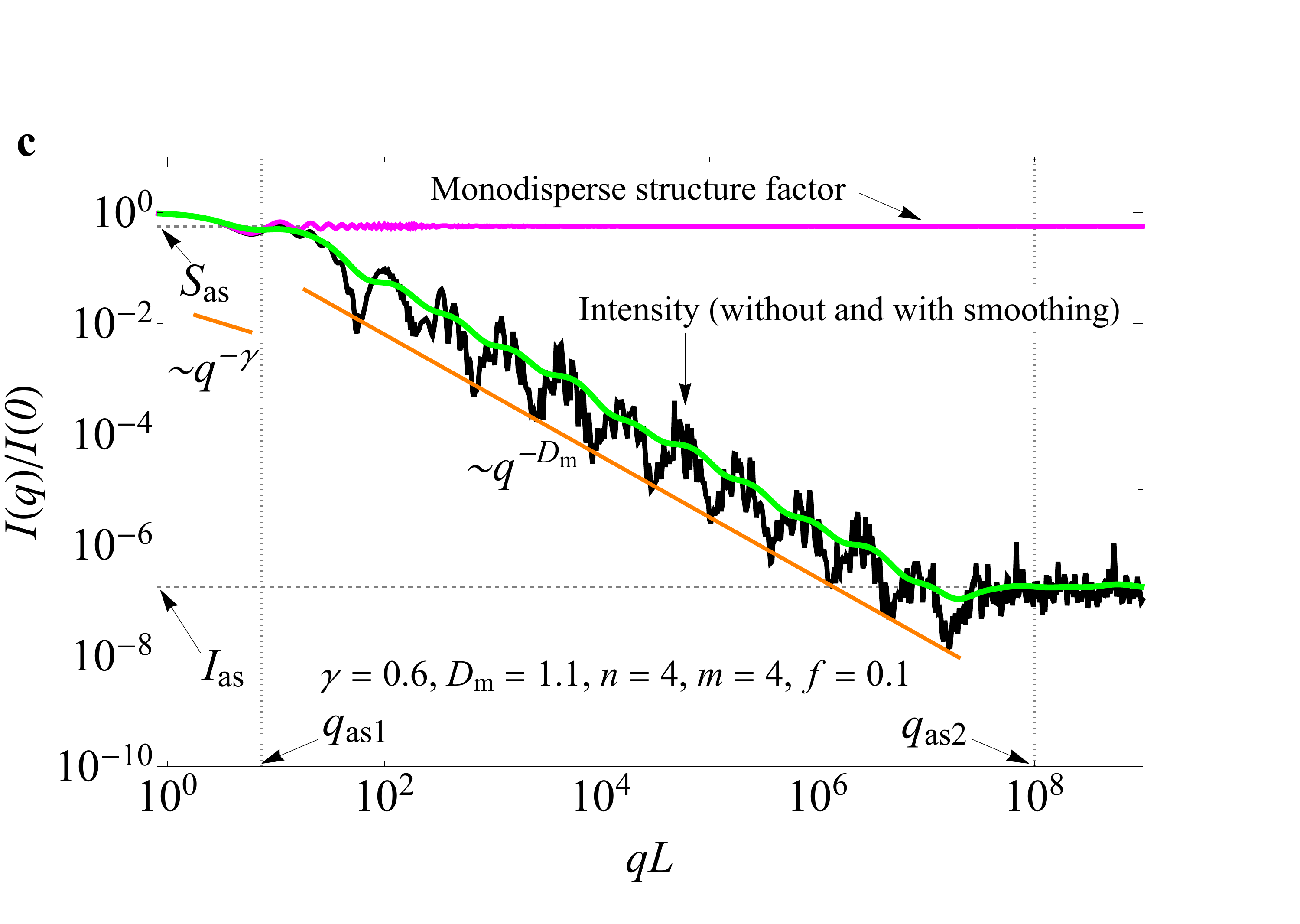}}
\caption{\label{fig:fig3} The normalized total (black) and structure-factor (magenta) intensities, given by Eq.~\eqref{I_S}, vs. the momentum transfer (in units of $1/L$) for the structure shown in Fig.~\ref{fig:discretemodel}. $L$ is the overall size of the entire structure. Smoothed intensity (\ref{eq:polyI}) [shown in green (light gray)] is obtained with Eq.~\eqref{eq:DN}. ({a}) The regime ${\gamma} > D_{\mathrm{m}}$ and $2 D_{\mathrm{m}} - {\gamma} > 0$. ({b}) ${\gamma} > D_{\mathrm{m}}$ and $2D_{\mathrm{m}} - {\gamma} = 0$. ({c}) ${\gamma} < D_{\mathrm{m}}$. Vertical dotted lines indicate the borders of the ranges with different exponents (see the discussion in the text). Dashed horizontal lines show the asymptotics $S_{\mathrm{as}}$ (\ref{Ias}) and $I_{\mathrm{as}}\simeq 1/N_{\mathrm{tot}}$. Here $N_{\mathrm{tot}}$ is the total number of scattering points in the entire structure.}
\end{figure}

Figure \ref{fig:fig3} represents the scattering curves for the model of the discrete power-law distribution of CMF (see Fig.~\ref{fig:discretemodel}). All three main cases are considered: $\gamma > D_{\mathrm{m}}$ and $2D_{\mathrm{m}} - {\gamma} > 0$ (Fig.~\ref{fig:fig3}{a}), ${\gamma} > D_{\mathrm{m}}$ and $2D_{\mathrm{m}} - {\gamma} = 0$ (Fig.~\ref{fig:fig3}{b}), and ${\gamma} <  D_{\mathrm{m}}$ (Fig.~\ref{fig:fig3}{c}). In all subfigures, the monodisperse structure factors $S(q)$ [Eqs.~(\ref{FStotal}) and \eqref{I_S}] are given in magenta, the monodisperse scattering intensities $I(q)$ [Eqs.~(\ref{Ftotal}) and \eqref{I_S}] in black, and the smoothed intensities $I_{\mathrm{sm}}(q)$ [Eq.~\eqref{eq:polyI}] in green (light gray).

The smoothed curve of $I(q)$ in Fig.~\ref{fig:fig3}{a} is characterized  by a succession of three power-law regimes. For $2\pi \lesssim qL \lesssim q_{\mathrm{as1}} L\equiv S_{\mathrm{as}}^{-1/{\gamma}}$, the scattering exponent is $\alpha = D_{\mathrm{tot}} ={\gamma}$ in accordance with Eq.~(\ref{eq:gamM}), where the upper border of this fractal range is given by Eq.~\eqref{eq:qas}. For $q_{\mathrm{as1}} \lesssim q \lesssim q_{\mathrm{m}}$ the scattering exponent is described by Martin's formula (\ref{eq:gamMmar}) $\alpha = 2D_{\mathrm{m}} - {\gamma}$, where the upper border of this fractal range is $q_{\mathrm{m}} \simeq 2\pi/l_{n}$, where $l_{n}=Lf(1-2\beta)\beta^{n}$ [with $\beta = (1/4)^{1/{\gamma}}$] is the size of the smallest block (see the discussion in Sec.~\ref{sec:modeldiscrete}). Since this block consists of the $m$th iteration of CMF, the curve decreases further as the mass-fractal intensity does \cite{cherny11}: when $q_{\mathrm{m}} \lesssim q \lesssim q_{\mathrm{as2}}$, we have $\alpha = D_{\mathrm{m}}$, where $q_{\mathrm{as2}} \simeq q_{\mathrm{m}}/\beta_{\mathrm{m}}^{m}$. Finally, all the correlations decay, and for $q \gtrsim q_{\mathrm{as2}}$ the asymptotics $I_{\mathrm{as}}\simeq 1/N_{\mathrm{tot}}$ is attained, where $N_{\mathrm{tot}}$ is the total number of scattering points in the entire structure. As expected, the total and structure-factor intensities coincide up to $q\simeq q_{\mathrm{as1}}$ as long as the spatial correlations between different fractals play a role.  This region is then followed by the asymptotics (\ref{Ias}) with a very good accuracy.

For the control parameters of Fig.~\ref{fig:fig3}{b}, the both intensities $I(q)$ and $S(q)$ are very similar to the previous case. The main difference is that for $q_{\mathrm{as1}} \lesssim q \lesssim q_{\mathrm{m}}$, the smoothed curve is almost constant. This is because Martin's relation (\ref{eq:gamMmar} ) yields $\alpha=2D_{\mathrm{m}} - {\gamma} = 0$.  The length of this range is  controlled by the parameter $f$.

Figure \ref{fig:fig3}{c} shows the scattering curves in the regime ${\gamma} < D_{\mathrm{m}}$.
The power-law decay with $\alpha = {\gamma}$ is very short and practically invisible.
We observe that the exponent of the smoothed intensity $I(q)$ is equal to $\alpha =D_{\mathrm{tot}}=D_{\mathrm{m}}$ in accordance with the both formulas (\ref{eq:gamM}) and (\ref{eq:gamMmar}). In this regime, the main contribution to the intensity comes from CMF of the same size, and correlations between CMF with different sizes are negligible. Note that the maximal CMF iteration in the central block amounts to $m + p(0) = 11$ with $p(0) = 7$ by Eq.~\eqref{pi}. This is confirmed by the presence of 11 pronounced minima of $I(q)$, superimposed on the power-law decay.

\section{
Classical Apollonian gasket}
\label{sec:cont1}

\subsection{Model}
\label{sec:appolo}

\begin{figure}[!tb]
\centerline{\includegraphics[width=0.97\columnwidth]{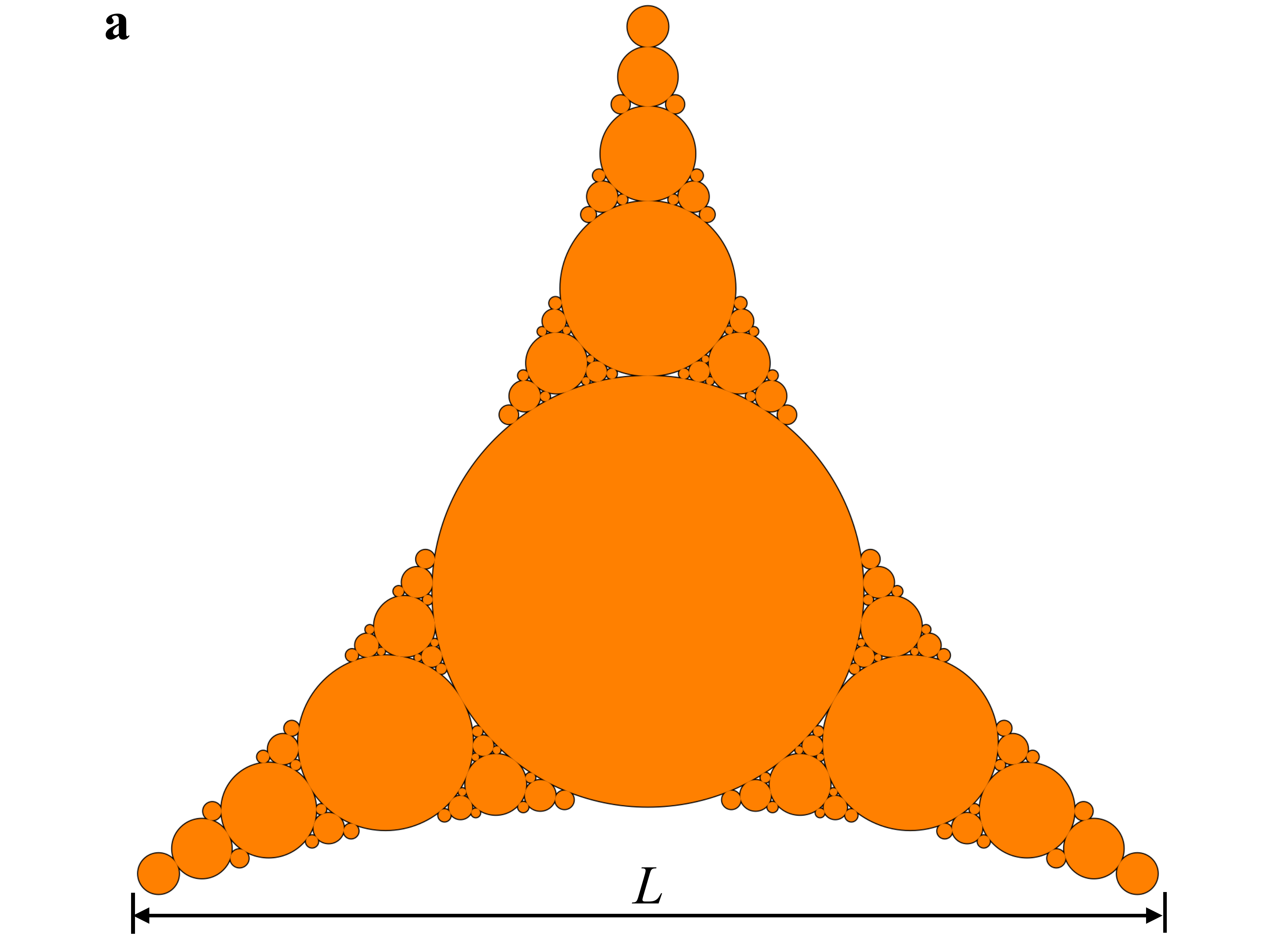}}
\vspace{0.3cm}
\centerline{\includegraphics[width=0.97\columnwidth]{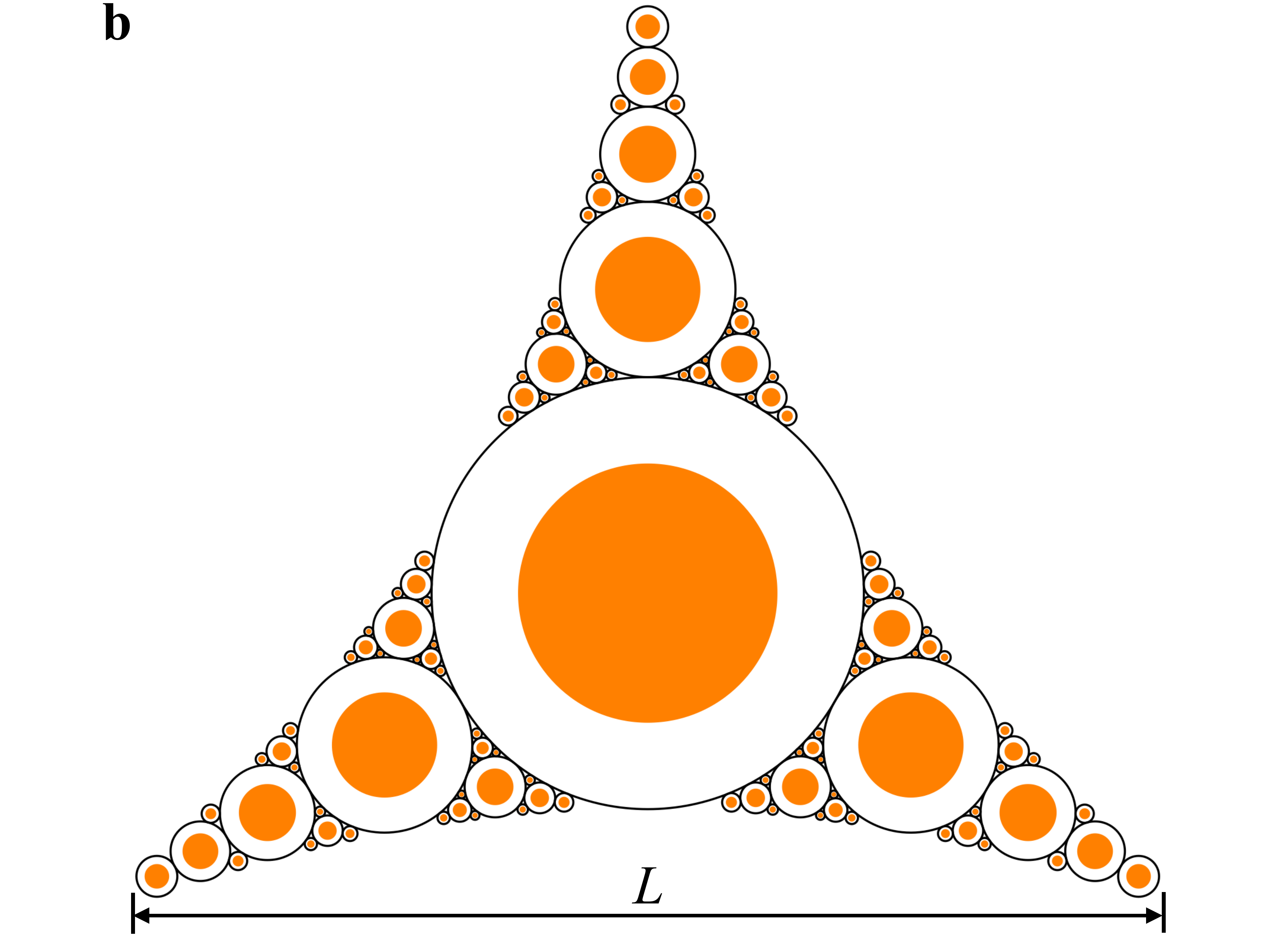}}
\caption{\label{fig:fig4} The 2d Apollonian gasket (orange) consisting from 121 disks. ({a}) The scaling factor $f = 1$. ({b}) $f = 0.6$. Black circles are added for better visualisation.}
\end{figure}

Continuous power-law distributions are simulated by the 2d Apollonian gaskets (AG). The construction of AG starts from a three initial disks, each one tangent to the other two, and filling the space between them with disks of smaller radii such that each is tangent to another three. Fig.~\ref{fig:fig4}a shows such a construction for 121 disks. The central disk has the radius $R_{0}$ and is tangent to all three initial disks (not shown here since they do not belong to AG, but they can be imagined as being situated at the left, right, and bottom parts of the AG, respectively, as indicated by their arcs). The next three smaller disks (of radii $R_{1}$) are placed such that they are tangent to the disk of radius $R_0$ and two of the initial three disks. Repeating this procedure infinitely leads to complete filling of the space between the three initial disks. The resulting set of disk borders (circles) forms a fractal with with fractal dimension $D_{\mathrm{s}}= 1.31\ldots$ \cite{PhysRevE.81.061403}. Besides, the distribution of disk sizes obeys a power law with the exponent ${\gamma}=D_{\mathrm{s}}$.

It is convenient to construct a modification of the AG. Let us keep the positions of the disk centers unchanged and scale their radii by a factor $f<1$. A typical configuration is shown in Fig.~\ref{fig:fig4}{b} for $f = 0.6$. We denote the overall size of the AG as $L$. Note that $L$ is independent of the scaling factor $f$.

For instructive purposes, we adopt a simplified model of mass-fractal structure inside each disk of AG. We assume that only the mass of a disk of radius $R_{j}$ is proportional to $R_{j}^{D_{\mathrm{m}}}$ in accordance with the mass-radius relation for fractals \cite{Gouyet1996PhysicsStructures}. Below in section \ref{sec:appoloCMF}, we consider a more realistic model.

\subsection{Scattering properties}

We calculate the scattering amplitude from the AG of Fig.~\ref{fig:fig4} with the methods of Refs.~\cite{Cherny2010ScatteringFractals,cherny11}. It is assumed that the disks are composed from a mass fractal of dimension $D_{\mathrm{m}}$, and hence the amplitude $A_{j}(q)$ of a disk of radius $R_{j}$ is proportional to  $R_{j}^{D_{\mathrm{m}}}$.  Then the normalized scattering amplitude is given by
\begin{align}\label{FAG}
F(\bm{q})=\frac{\sum_{j}A_{j}(q)e^{i\bm{q}\cdot\bm{r}_{j}}}{\sum_{j}A_{j}(q)},\quad A_{j}(q)=R_{j}^{D_{\mathrm{m}}}F_{0}(R_{j}q),
\end{align}
where vectors $\bm{r}_{j}$ are positions of the disk centers. Here the normalized scattering amplitude of the disk of unit radius is given by $F_{0}(q) = 2J_{1}(q)/q$ with $J_{1}(q)$ being the Bessel function of the first kind. The structure-factor amplitude $G(\bm{q})$ is calculated with the same equation (\ref{FAG}) but with $F_{0}=1$ as if the scatterers were point-like objects. The normalized SAS intensity and structure factor are obtained by means of relations (\ref{I_S}).

For sufficiently large momenta, only the diagonal terms survive due to the randomness of the phases in the exponential. As a result, we obtain the asymptotics of the structure factor of AG
\begin{equation}
\label{eq:lasAG}
S_{\mathrm{as}}=\frac{\sum_{i}R_{i}^{2 D_{\mathrm{m}}}}{\left(\sum_{i}R_{i}^{D_{\mathrm{m}}}\right)^2}.
\end{equation}

\begin{figure}[!tb]
\centerline{\includegraphics[width=\columnwidth,trim={0 1cm 2.5cm 2cm}]{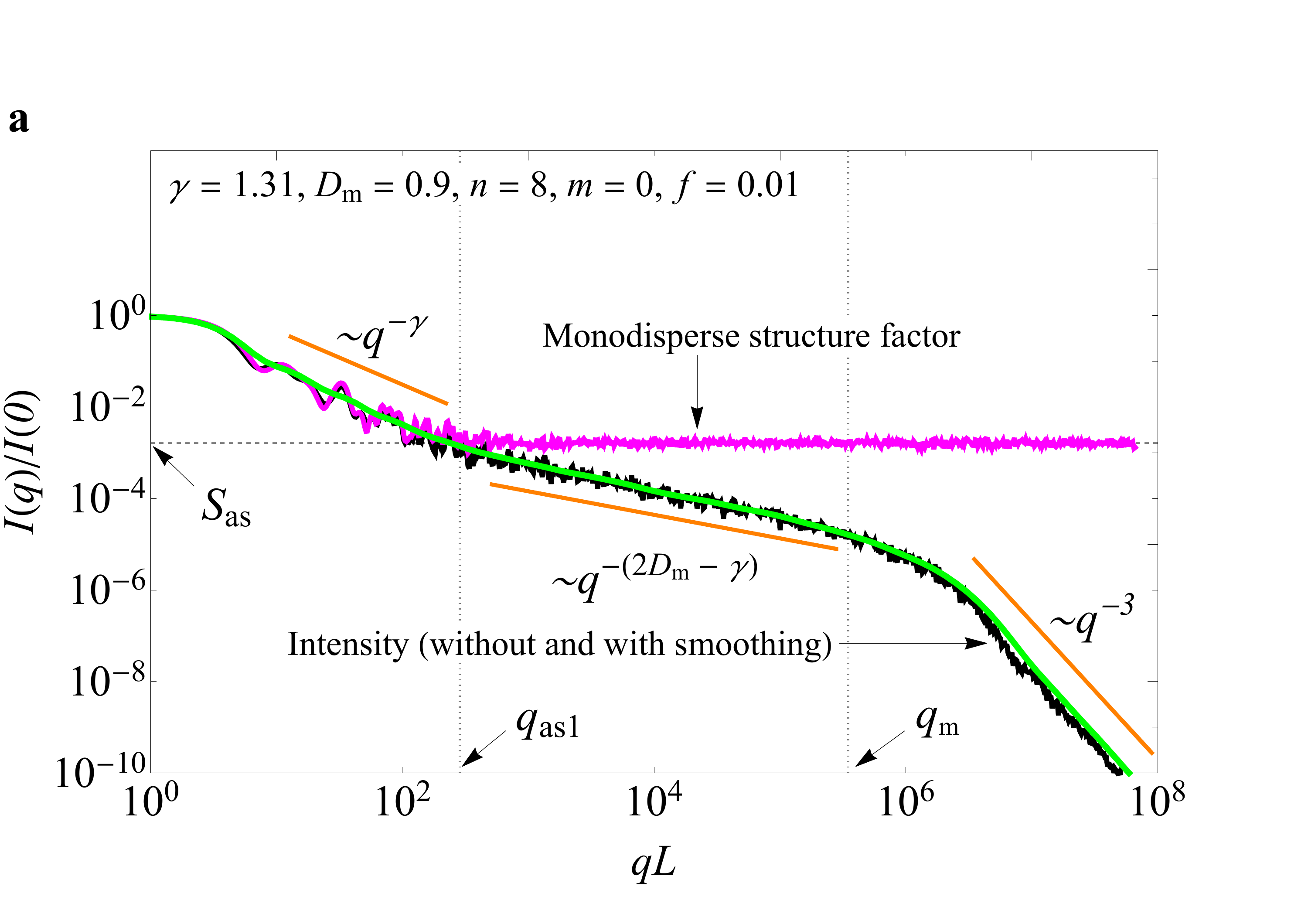}}
\centerline{\includegraphics[width=\columnwidth,trim={0 1cm 2.5cm 2cm}]{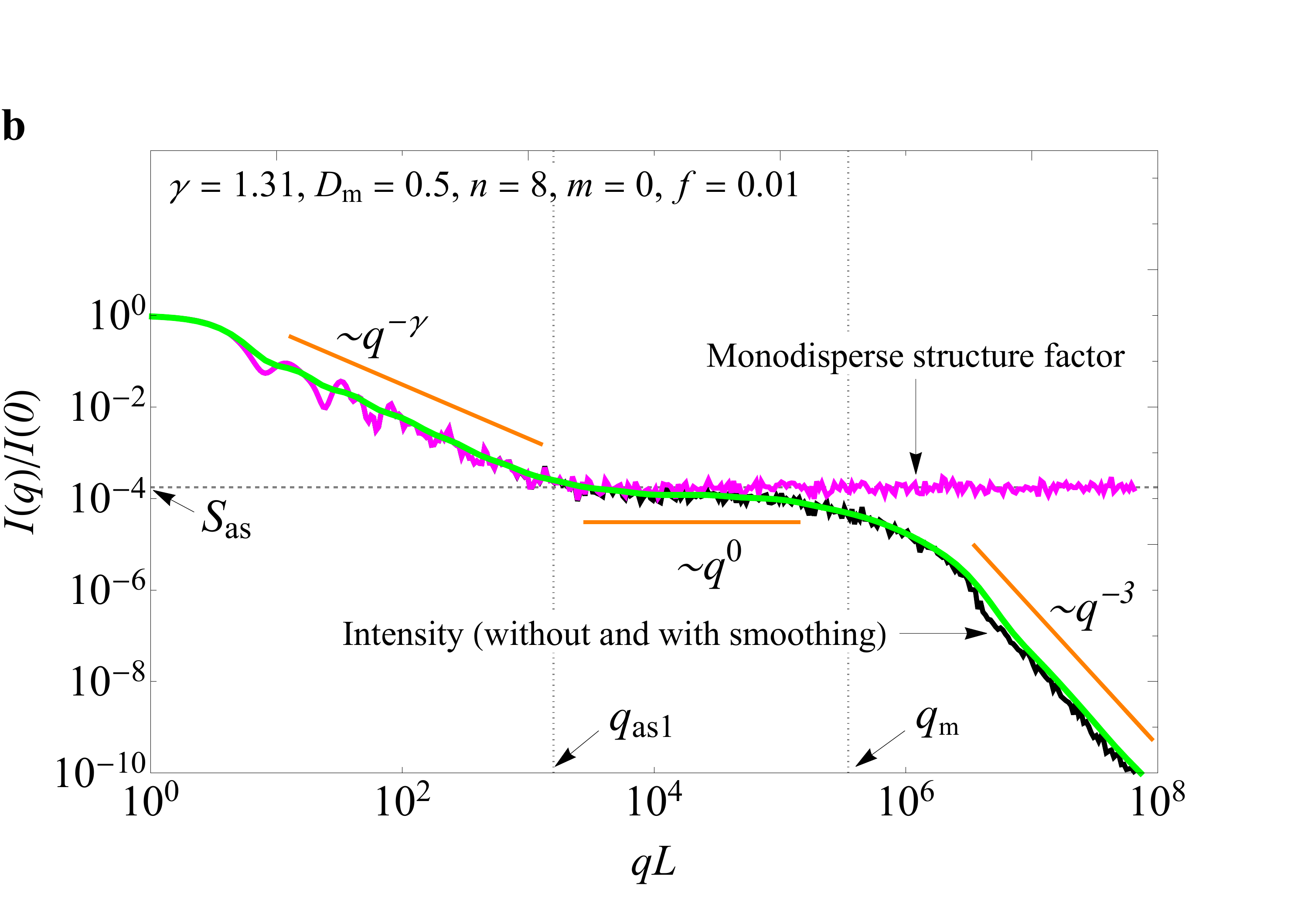}}
\caption{\label{fig:fig5} The normalized total intensity (black), smoothed intensity [green (light gray)], and structure factor (magenta) vs. momentum transfer (in units of the inverse overall size $1/L$) for the structure shown in Fig.~\ref{fig:fig4}. ({a}) ${\gamma} > D_{\mathrm{m}}$ and $2 D_{\mathrm{m}} - {\gamma} > 0$. ({b}) ${\gamma} > D_{\mathrm{m}}$ and $2D_{\mathrm{m}} - {\gamma} < 0$.  Vertical dotted lines mark the regions with different power-law exponents (see the main text for details). Dashed horizontal line represents the asymptotic values $S_{\mathrm{as}}$ (\ref{eq:lasAG}).}
 \end{figure}

Figure \ref{fig:fig5} shows the scattering curves when ${\gamma}> D_{\mathrm{m}}$ and $2D_{\mathrm{m}} -{\gamma} > 0$ (Fig.~\ref{fig:fig5}a) and $2D_{m}-{\gamma} < 0$ (Fig.~\ref{fig:fig5}b). The color coding is the same as for Fig.~\ref{fig:fig3}, and the total number of disks is equal to $\sum_{i=0}^{n} 3^{i} = 9841$ for the number of iterations of AG $n=8$.

The scattering intensities are very similar to those of Fig.~\ref{fig:fig3}. For $2\pi/L \lesssim q \lesssim q_{\mathrm{as1}}$, we obtain again $\alpha = D_{\mathrm{tot}}={\gamma}$ for ${\gamma}> D_{\mathrm{m}}$ by Eq.~(\ref{eq:gamM}). Here, $q_{\mathrm{as1}}$ is given again by Eq.~\eqref{eq:qas} with asymptotic values $S_{\mathrm{as}}$ (\ref{eq:lasAG}). It follows from these equations that the length of this region increases with increasing the mass-fractal dimension $D_{\mathrm{m}}$.  For $q_{\mathrm{as1}} \lesssim q \lesssim q_{\mathrm{m}}$, we also recover $\alpha = 2D_{\mathrm{m}} - {\gamma}$, where $q_{\mathrm{m}} = 2\pi / (f \beta^{n})$ and $\beta = (1/3)^{1/{\gamma}}$. For $q \gtrsim q_{\mathrm{m}}$, as opposed to scattering from discrete power-law distribution in Fig.~\ref{fig:fig3}, the power-law decay obeys the Porod law $\alpha = 3$ in two dimensions. This is because disks are regular non-fractal structures.

The structure factor describes the scattering from point-like objects weighted with the mass $R_{j}^{D_{\mathrm{m}}}$ of each disk, and it still decays with the exponent $\gamma$.  We conclude that the predicted exponent $\gamma$ [in accordance with Eq.~(\ref{eq:gamM})] arises due to spatial correlations between different mass fractals, and the internal structure of fractals does not play a role. For $q\gtrsim q_{\mathrm{as1}}$ the structure factor does not change anymore, which implies that these correlations completely decay. Then the exponent $2D_{\mathrm{m}} - {\gamma}$ appears as a sum of the intensities of the mass fractals in argeement with Martin's approach \cite{Martin1987ScatteringFractals,Schmidtrev91}.

\section{Apollonian gaskets consisting of Cantor mass fractals}
\label{sec:cont2}

\subsection{Model}
\label{sec:appoloCMF}

\begin{figure}[!tb]
\centerline{\includegraphics[width=0.95\columnwidth]{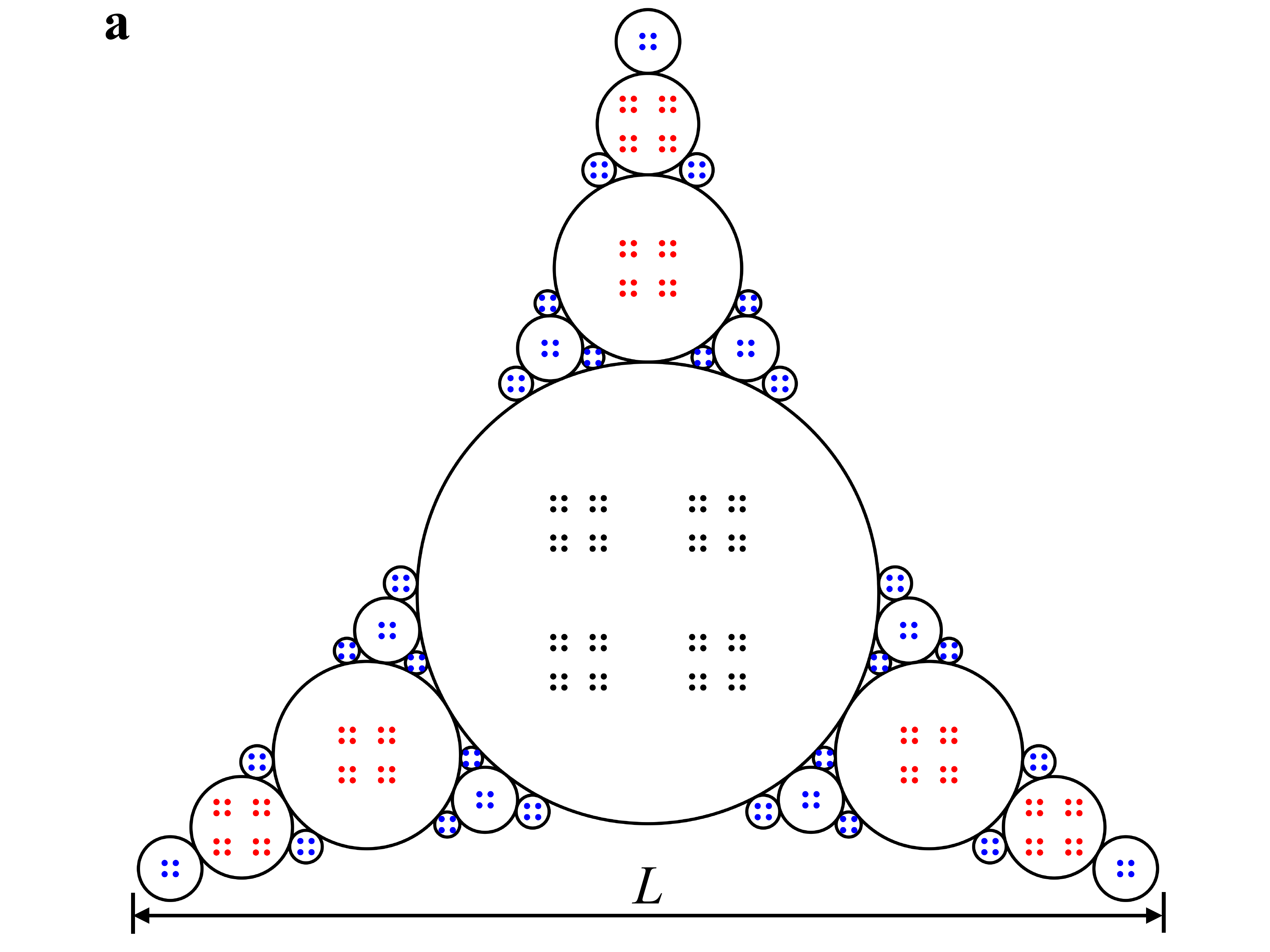}}
\vspace{0.3cm}
\centerline{\includegraphics[width=0.95\columnwidth]{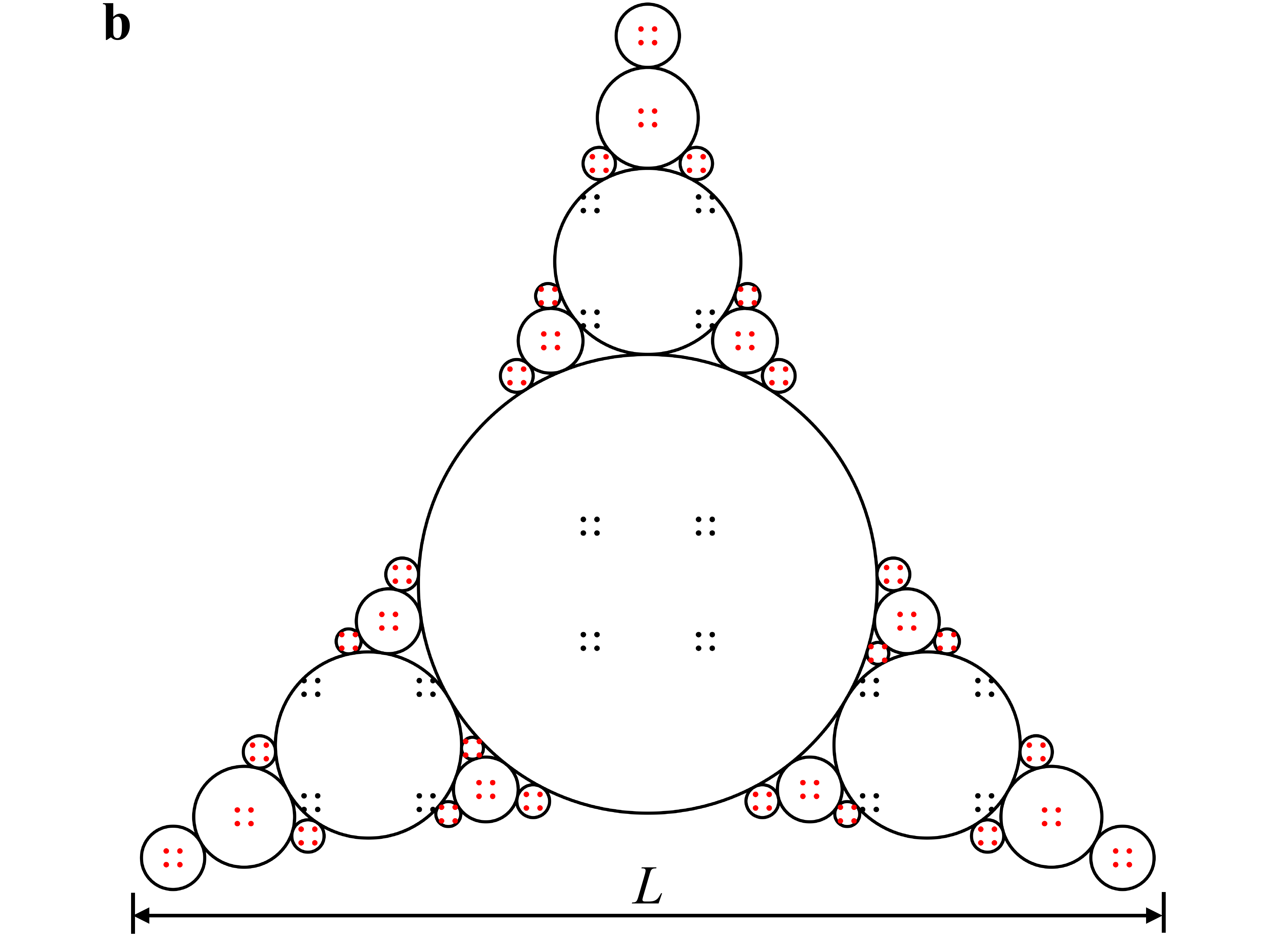}}
\caption{\label{fig:fig6} The model based on the AG shown in Fig.~\ref{fig:fig4}a: the disks are replaced by CMFs. The first 40 fractals are represented. The circles are imaginary and shown for better visualization. ({a}) $D_{\mathrm{m}} = 0.9$, for which  $2 D_{\mathrm{m}} - {\gamma} > 0$. The fractal iteration number inside the smallest AG disk is chosen to be $m=1$ (blue disks). Higher iterations of CMF $m=2$ and $m=3$ are given in red and black, respectively. ({b}) $D_{\mathrm{m}} = 0.65$, for which  $2 D_{\mathrm{m}} - {\gamma} = 0$. The fractal iteration number inside the smallest AG disk is chosen to be $m=1$ (red dots). The second iteration of CMF is indicated with black dots.}
\end{figure}

In the previous section, we did not specify a model of mass fractal inside the disks of AG. In this section, the approach is model-dependent:  we replace the disks by CMFs of dimension $D_{\mathrm{m}}$ (see Fig.~\ref{fig:fig6}). In turn, the CMFs are composed of point-like objects as for the discrete power-law distribution of Sec.~\ref{sec:models}. The smallest disks contain the smallest mass-fractal iteration $m$ of the same size ($m=1$ in the figure). Further, the maximal possible iterations more than $m$ are inscribed into the circles as long as their radii grow.

Specifically, Fig.~\ref{fig:fig6} shows the model for two sets of parameters at $f = 1$ and with CMF placed inside the first 40 disks of AG. Figure \ref{fig:fig6}{a} corresponds to the case $2D_{\mathrm{m}}-{\gamma} > 0$, where $D_{\mathrm{m}} = 0.9$. The construction is as follows: first, we consider CMF at various iterations $m$, with $m = 1, \cdots, m_{\mathrm{max}}$ (here $m_{\mathrm{max}} = 4$). The diagonal $d_1$ of CMF at $m = 1$ is set to be equal to $2r_{\mathrm{min}}$, where $r_{\mathrm{min}}$ is the smallest radius for the set of considered disks. Then, the diagonals $d_2, d_3$ and $d_4$ of CMF at $m=2, m=3$, and respectively at $m = 4$ are calculated using a bottom-up approach, as described in Sec.~\ref{sec:modeldiscrete}. This gives $d_2 = \beta_{\mathrm{m}}^{-1} d_1 > d_1$ since $0 < \beta_{\mathrm{m}} < 1/2$, $d_3 = \beta_{\mathrm{m}}^{-2} d_1 > d_2$, and respectively $d_4 = \beta_{\mathrm{m}}^{-4} d_1 > d_3$. The CMF of size $d_1$ is placed in all AG disks whose diameters are smaller than $d_2$ (blue disks). The CMF of size $d_2$ is placed in all AG disks whose diameters are greater than or equal to $d_2$ but smaller than $d_3$ (red disks). Finally, the CMF at $m = 3$ is placed in all AG disks whose diameters are greater than or equal to $d_3$ but smaller than $d_4$ (black disks). The same procedure is used in Fig.~\ref{fig:fig6}{b} when $2D_{\mathrm{m}} - {\gamma} = 0$ with $D_{\mathrm{m}} = 0.65$. Here the smallest CMF iteration $m = 1$ is shown in red, while the iteration $m = 2$ is depicted in black. Note that the maximal number of CMF iterations $m_{\mathrm{max}}$ depends on $D_{\mathrm{m}}$. We emphasize that the circles are imaginary and serve only as a delimiter of the region occupied by AG disks.

\begin{figure}[!tb]
\centerline{\includegraphics[width=\columnwidth,trim={0 1cm 2.5cm 2cm}]{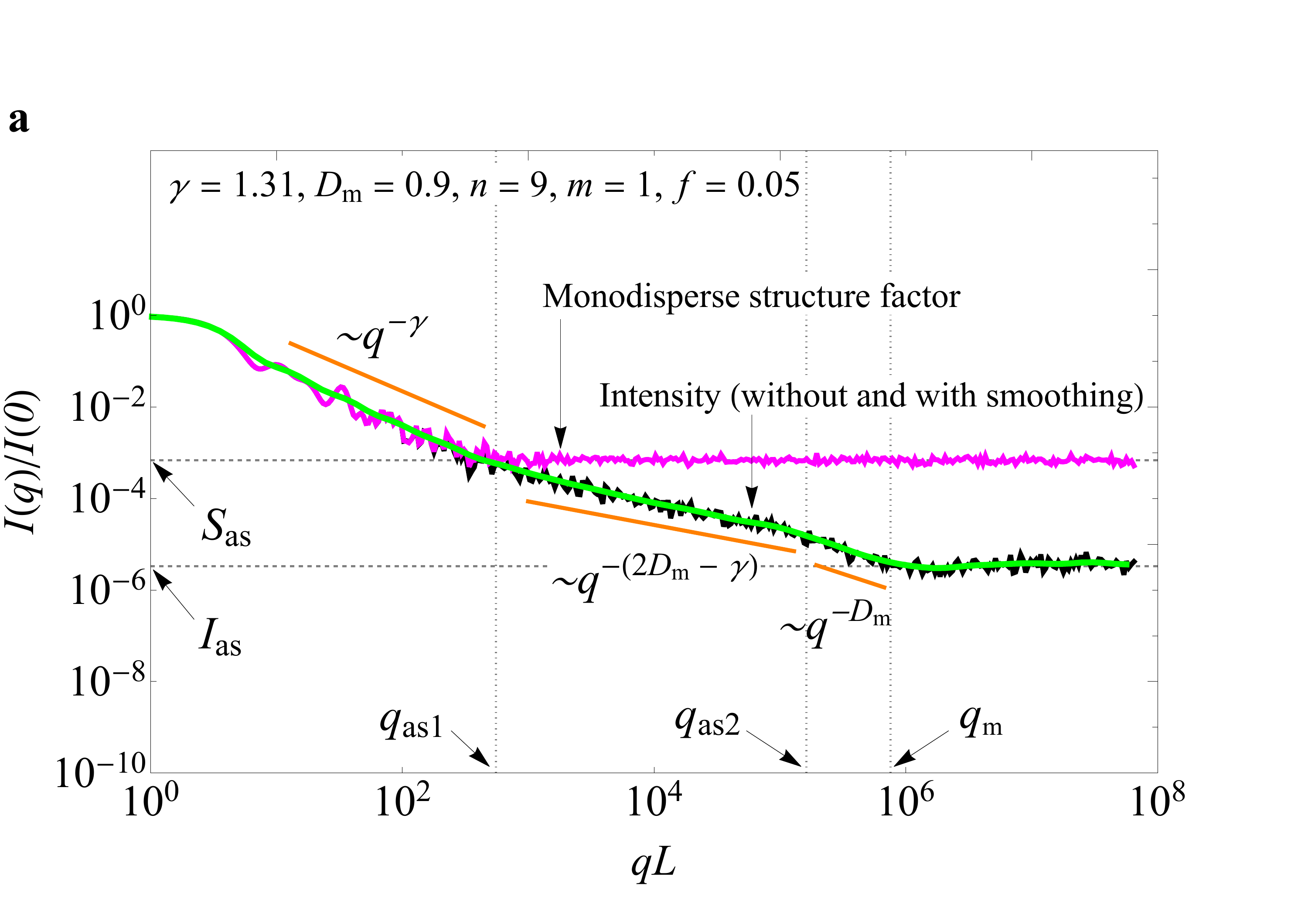}}
\centerline{\includegraphics[width=\columnwidth,trim={0 1cm 2.5cm 2cm}]{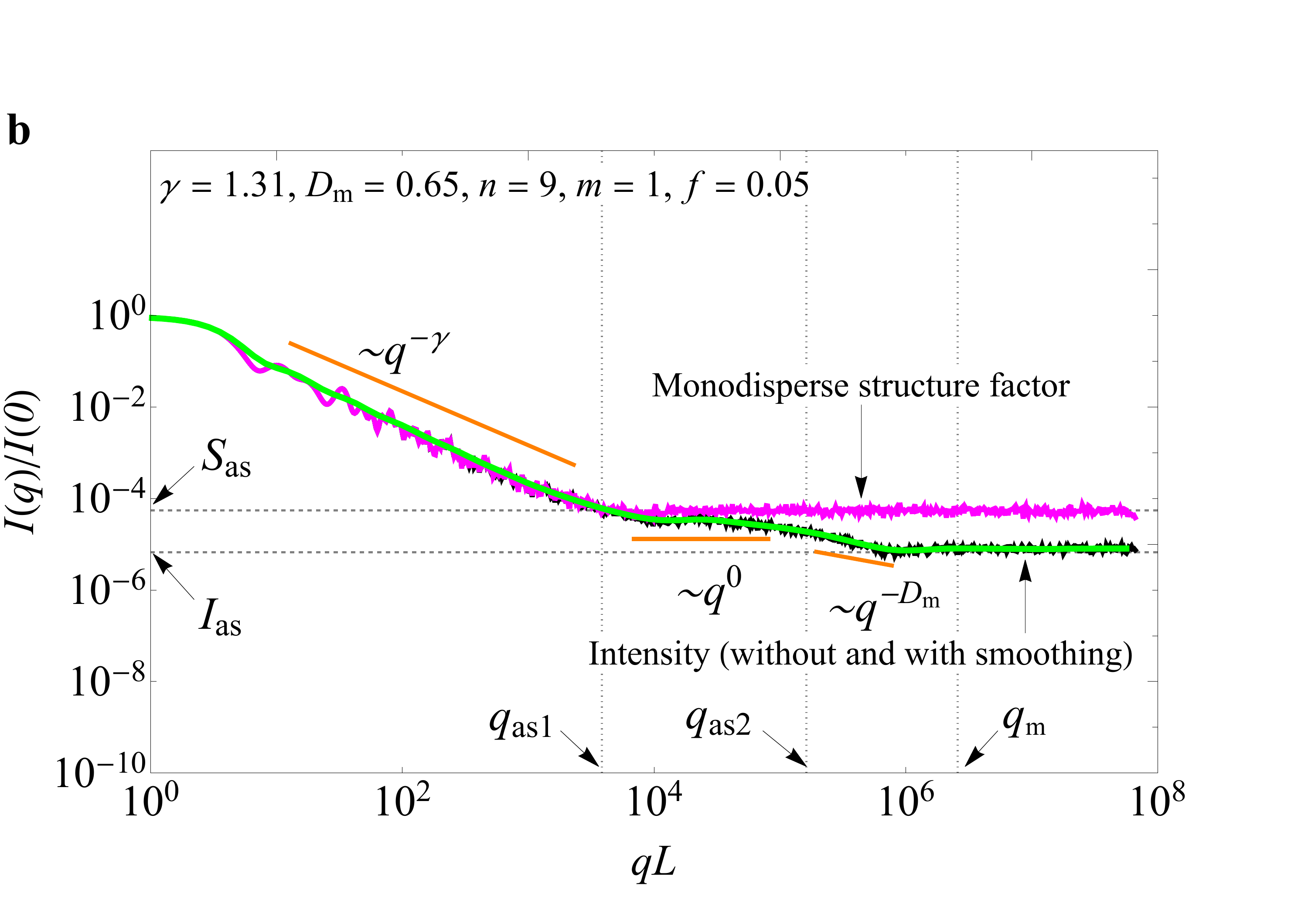}}
\caption{\label{fig:fig7} The scattering curves for the structure shown in Fig.~\ref{fig:fig6}. The notations are the same as in Fig.~\ref{fig:fig5}. ({a}) ${\gamma} > D_{\mathrm{m}}$ and $2 D_{\mathrm{m}} - {\gamma} > 0$.  ({b}) ${\gamma} > D_{\mathrm{m}}$ and $2D_{\mathrm{m}} - {\gamma} < 0$. The lower dashed horizontal line is $I_{\mathrm{as}} = 1/N_{\mathrm{tot}}$, where $N_{\mathrm{tot}}$ is the total number of scattering points.}
\end{figure}

\subsection{Scattering properties}

The scattering amplitude is calculated by analogy with the previous section. It is given by Eq.~(\ref{FAG}) with the amplitudes of CMF of the appropriate iteration and scale, which are determined by the CMF constructions described in Sec.~\ref{sec:appoloCMF}. The total number of disks $\sum_{i=0}^{n} 3^{i} \equiv 29524$ for $n=8$ iterations of AG.

The behaviour of the scattering curves (Fig.~\ref{fig:fig7}) is similar to those in Fig.~\ref{fig:fig5}. As expected, instead of the Porod decay in Fig.~\ref{fig:fig5}, we observe the decay of CMF with $\alpha=D_{\mathrm{m}}$. The range of the mass-fractal behaviour is very short, since $m=1$. Due to the point-like structure of the entire construction, the intensity does not fall off to zero but tends to  $I_{\mathrm{as}} = 1/N_{\mathrm{tot}}$, where $N_{\mathrm{tot}}$ is the total number of scattering points.

\section{Dense random packing with a power-law size distribution}
\label{sec:rpl-disks}

The models suggested in the previous sections might seem somewhat artificial. In this section, we consider a more realistic model of dense random packing of disks obeying a power-law size distribution. Such distributions are often used in the literature to describe the structure of various systems such as colloids, biological systems, gases or granular materials \cite{torquito}. In particular, the degree of packing fraction plays an important role in the coalescence of concentrated high internal-phase-ratio emulsions \cite{kwok} and in the disorder-order phase transitions \cite{tor10}.

\subsection{Model}

\begin{figure}[h]
\centerline{\includegraphics[width=.9\columnwidth]{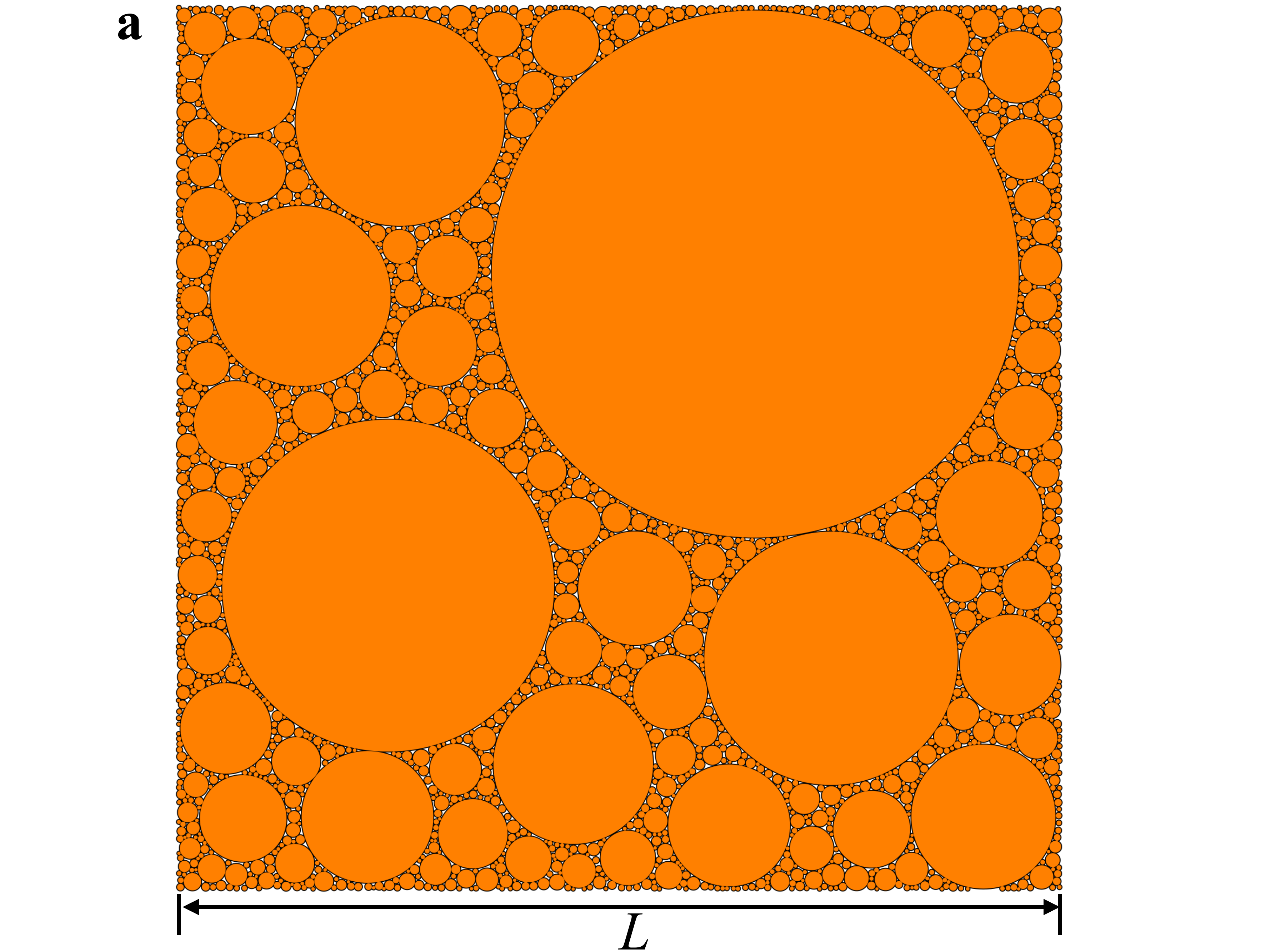}}
\vspace{0.3cm}
\centerline{\includegraphics[width=.9\columnwidth]{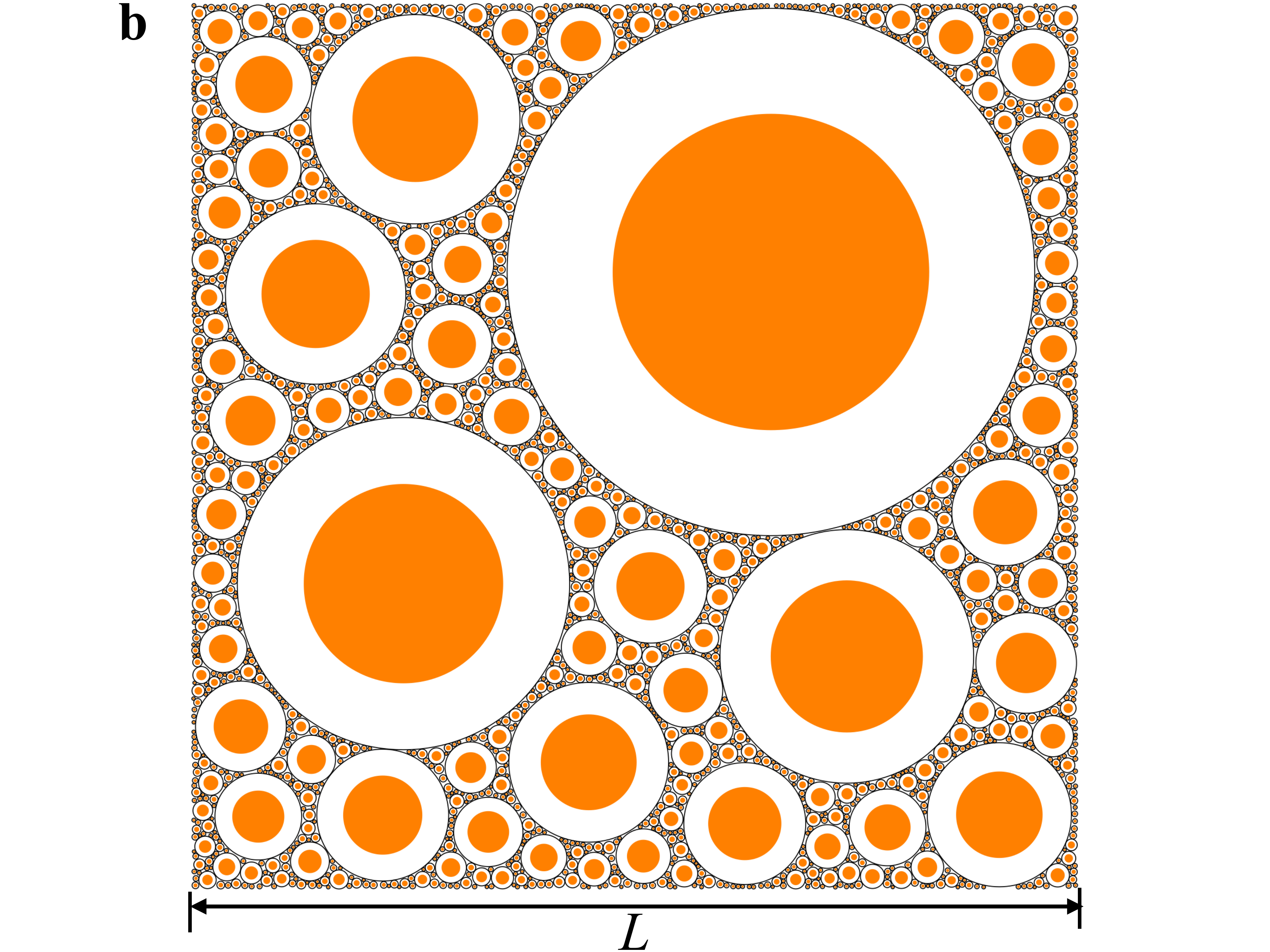}}
\caption{\label{fig:fig8}(a) Dense random packing of $N=2067$ disks inside a square. The disks radii follow the power-law distribution with the exponent $\gamma = 1.5$. The scaling factor $f = 1$. (b) The same diagram with $f = 0.6$. Black circles are added for better visualisation.}
 \end{figure}

We consider a set of $N$ non-overlapping disks randomly put into a square and with radii following a power-law distribution. The radii $R_{i}$ obey the inequalities $L/2 \geq R = R_1 > \cdots > R_N$, and $R_{i} = R\, i^{-1/D}$, where $i=1, \cdots, N$. Here $L$ is the edge of the square, and $R$ is the largest radius. Then, we put the center of the largest disk at a random position inside the square such that the entire disk is found inside the square. The same operation is repeated in turn for each of the remaining disks, which are all embedded in the remaining free space inside the square. By simple algebraic operations, we find the upper limit $R/L \simeq 0.329$ for which the algorithm can be applied  \cite{cherny22}. Figure~\ref{fig:fig8}a shows a configuration of disks with the packing fraction $0.937$ when $R/L \simeq 0.298$. By scaling the disks sizes by the factor $f = 0.6$ and keeping the positions of their centers unchanged, we obtain the configuration shown in Fig.~\ref{fig:fig8}b. As in the previous sections \ref{sec:cont1} and \ref{sec:cont2}, the disks are supposed to be replaced by mass fractals with fractal dimension $D_\mathrm{m}$. Thus we arrive at the model of a power-law polydispersity of mass fractals with random positions.

\subsection{Scattering properties}

\begin{figure}[h]
\centerline{\includegraphics[width=\columnwidth,trim={0 1cm 2.5cm 2cm}]{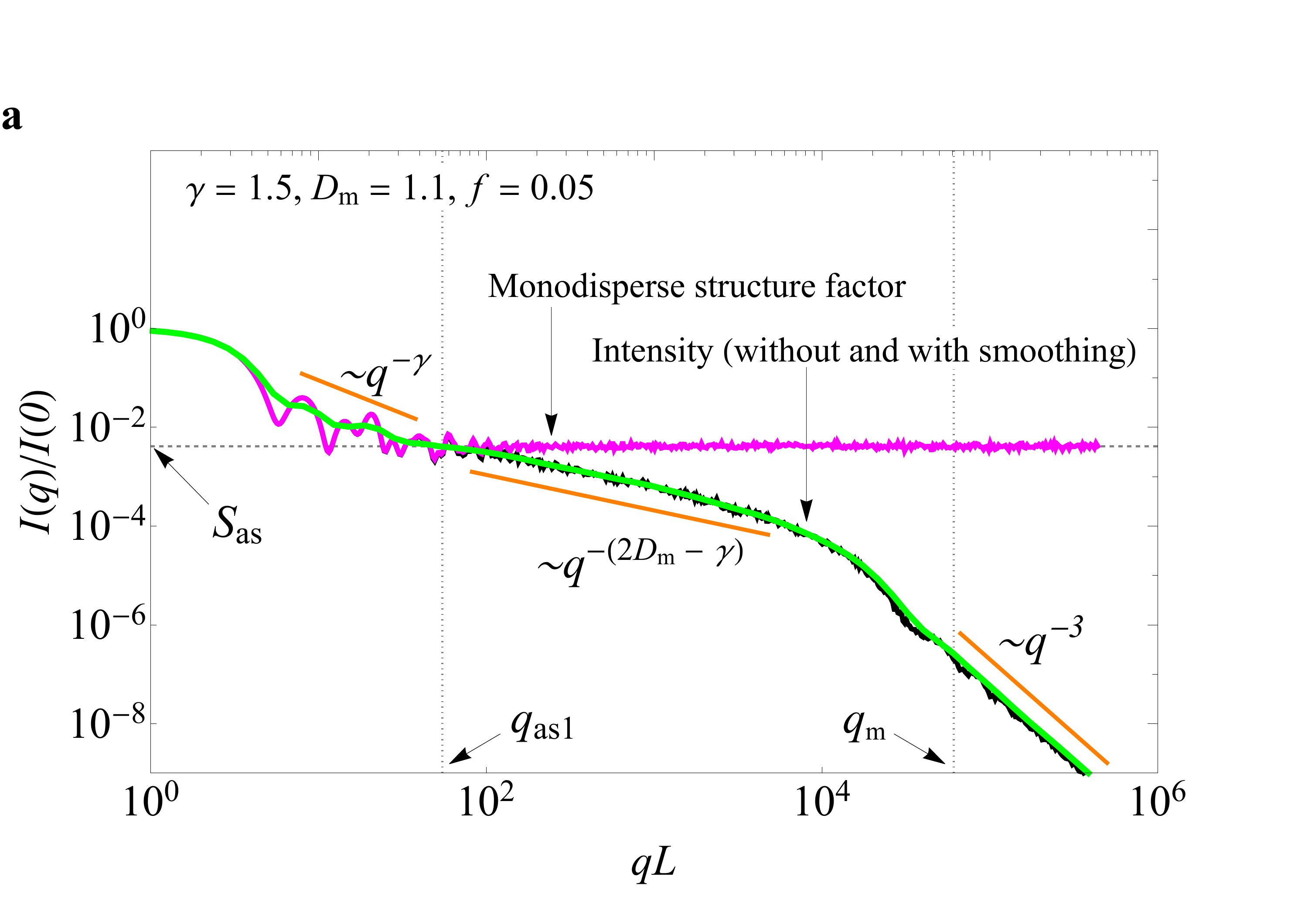}}
\centerline{\includegraphics[width=\columnwidth,trim={0 1cm 2.5cm 2cm}]{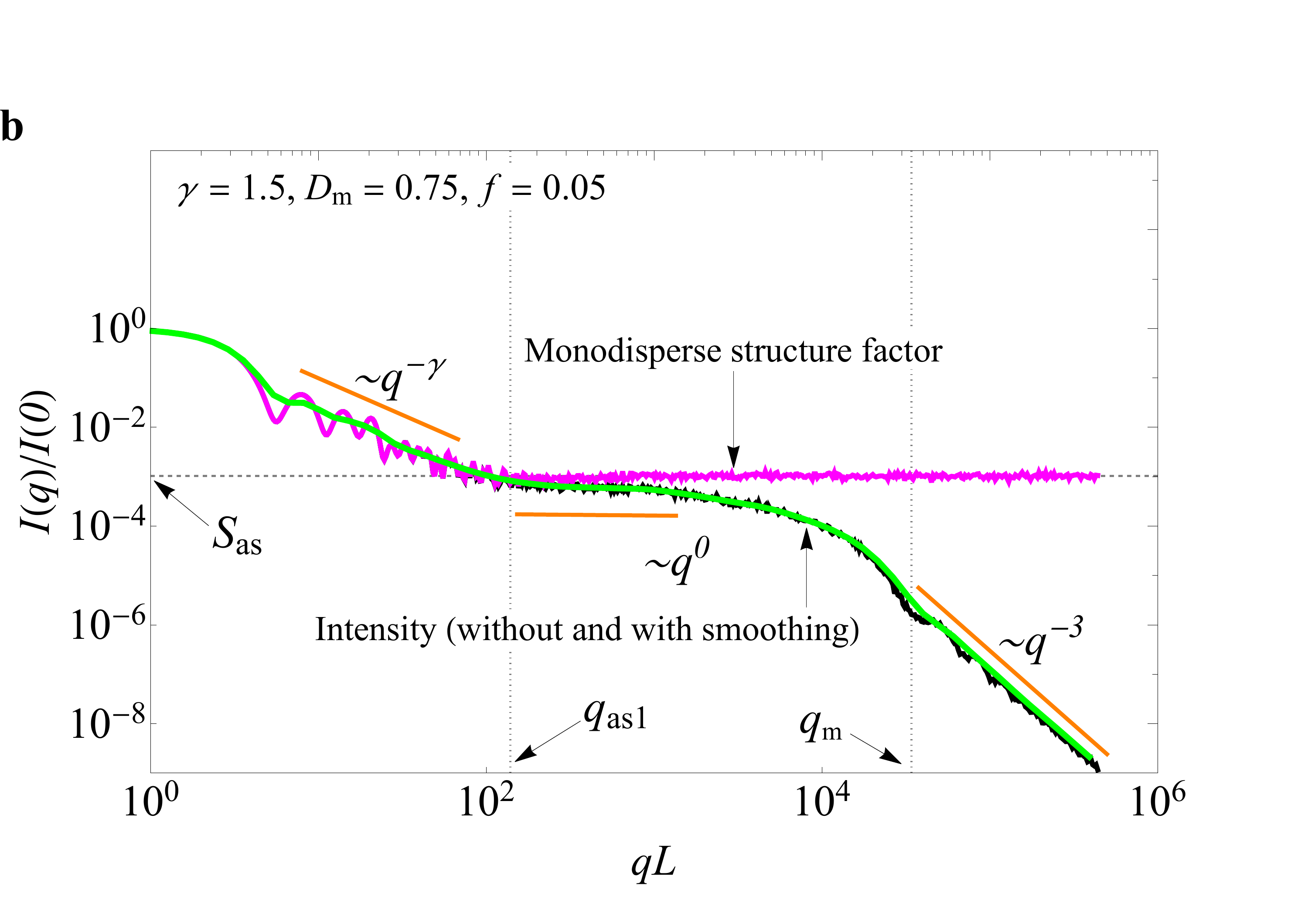}}
\caption{\label{fig:fig9} The normalized total intensity (black), smoothed intensity [green (light gray)], and structure factor (magenta) vs. momentum transfer (in units of the inverse overall size $1/L$) for the structure shown in Fig.~\ref{fig:fig8}. ({a}) ${\gamma} > D_{\mathrm{m}}$ and $2 D_{\mathrm{m}} - {\gamma} > 0$. ({b}) ${\gamma} > D_{\mathrm{m}}$ and $2D_{\mathrm{m}} - {\gamma} = 0$.  Vertical dotted lines mark the regions with different power-law exponents (see the main text for details). The dashed horizontal line represents the asymptotic values $S_{\mathrm{as}}$ (\ref{eq:lasAG}).}
\end{figure}

The scattering amplitudes are calculated also with Eq.~\eqref{FAG}, and the scattering intensity and the structure factor with Eq.~\eqref{I_S}. Figure~\ref{fig:fig9} shows the corresponding curves when $\gamma > D_{\mathrm{m}}$ and $2D_{\mathrm{m}} - \gamma > 0$ (Fig.~\ref{fig:fig9}a) and $2D_{\mathrm{m}} - \gamma = 0$ (Fig.~\ref{fig:fig9}b) for $N = 2067$ disks. In both cases, the scattering curves are qualitatively similar to that of AG shown in Fig.~\ref{fig:fig5}. We recover the exponents $\alpha = D_{\mathrm{tot}} = \gamma$ for $\gamma > D_{\mathrm{m}}$ and $2\pi / L \lesssim q \lesssim q_{\mathrm{as1}}$, $\alpha = 2D_{\mathrm{m}} - \gamma$ for $q_{\mathrm{as1}} \lesssim q \lesssim q_{\mathrm{m}}$, and $\alpha = 3$ for $q_{\mathrm{m}} \lesssim q$. Here, $q_{\mathrm{as1}}$ is given by the same Eq.~\eqref{eq:qas}, but $q_{\mathrm{m}}$ is estimated directly from the minimal distance $\delta$ between disks centers, i.e $q_{\mathrm{m}} \simeq 2\pi/\delta$.

\begin{figure}[h]
\centerline{\includegraphics[width=\columnwidth]{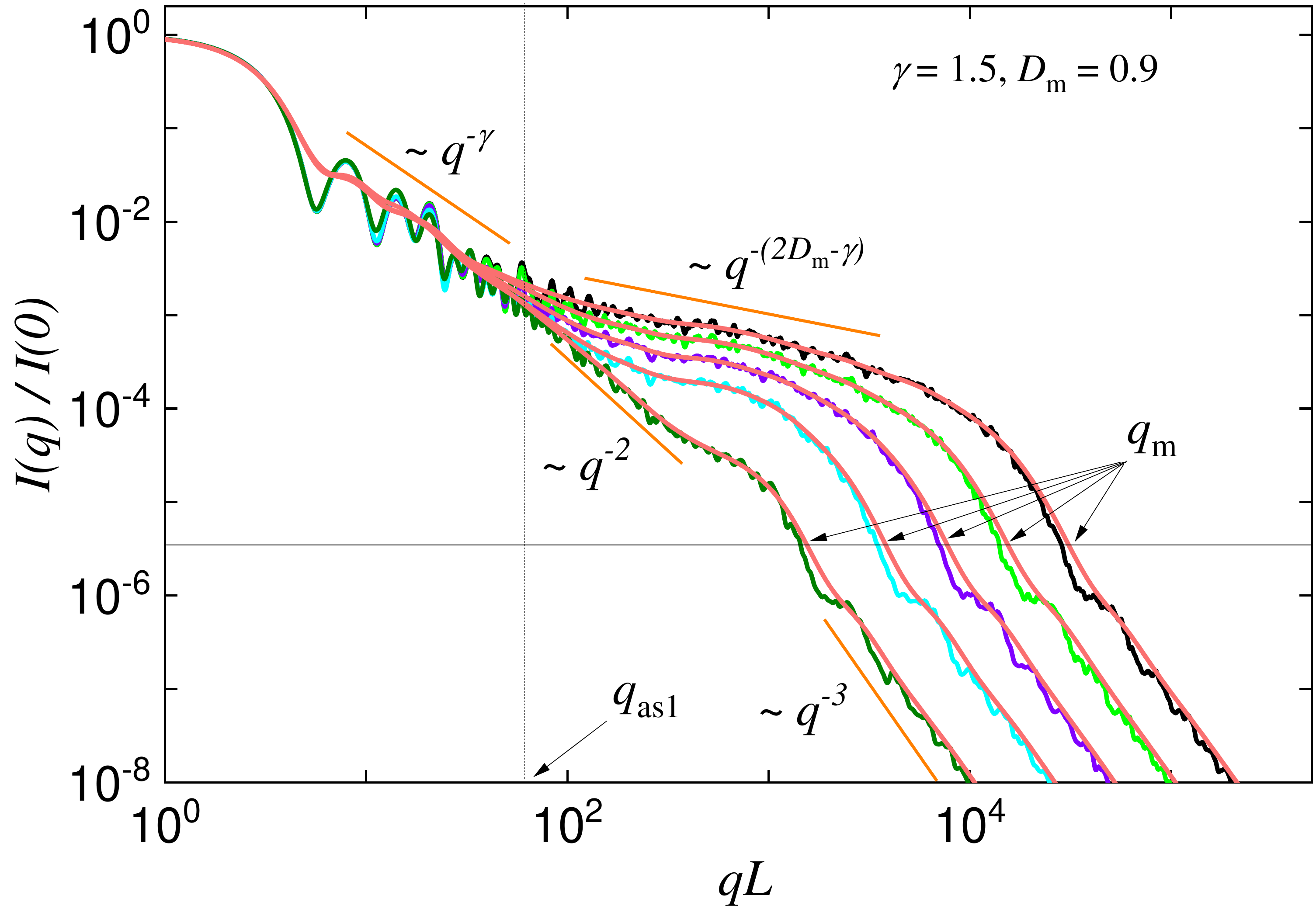}}
\caption{\label{fig:fig10} The normalized total intensities without and with smoothing at various values of parameter $f$ vs. momentum transfer (in units of the overall size $1/L$) when $2D_{\mathrm{m}} - \gamma > 0$. The relative variance $\sigma_{\mathrm{r}}$ for the smoothed curves is equal to $0.2$. From left to right: $f=1$, $f=0.4$, $f=0.2$, $f=0.1$ and $f=0.05$. The vertical dotted line marks the transition of intensities with exponents from $\alpha = \gamma$ to $\alpha = 2D_{\mathrm{m}}-\gamma$. The intersection of the continuous horizontal line at about 2.5$\times 10^{-6}$ with scattering curves is an estimation of the position of the transition point between intensities with exponents $\alpha = 2D_{\mathrm{m}} - \gamma$ and $\alpha = 3$.
}
\end{figure}

The scaling factor $f$ is directly related to the system porosity, but it does not change the distance correlations between disk positions.
We check whether the qualitative behaviour of scattering intensity remains unchanged for highly concentrated systems. Figure~\ref{fig:fig10} represents the normalized scattering intensities at $\gamma = 1.5$ and $D_{\mathrm{m}} = 0.9$ for various values of $f$. The results show that the slope $\alpha = \gamma$ in the region $q \lesssim q_{\mathrm{as1}}$ is kept unchanged for each value of $f$. However, the length of the region with $\alpha = 2D_{\mathrm{m}}- \gamma$ decreases, since the crossover point between intensities with exponents $\alpha = 2D_{\mathrm{m}} - \gamma$ and $\alpha = 3$ shifts to the left with increasing $f$.

Thus the behaviour predicted by our Eq.~\eqref{eq:gamM} is still visible for concentrated systems.
The smaller the scaling factor $f$, the more pronounced the behaviour with Martin's exponent (\ref{eq:gamMmar}).
Note that in the absence of scaling ($f = 1$), Martin's exponent $\alpha = 2D_{\mathrm{m}}-\gamma$ is surprisingly replaced by the exponent $\alpha = 2$.

\section{Dense random packing of non-overlapping cantor mass fractals}
\label{sec:rpl-CMF}

In the previous section, the mass-fractal nature of the disks was taken into consideration by choosing the appropriate weight $R_{j}^{D_{\mathrm{m}}}$ when calculating the total scattering amplitude [see Eq.~(\ref{FAG})]. Here we consider a specific ``microscopic'' model of the fractals by analogy with Sec.~\ref{sec:cont2}.

\subsection{Model}
\begin{figure}[tb!]
\centerline{\includegraphics[width=.9\columnwidth]{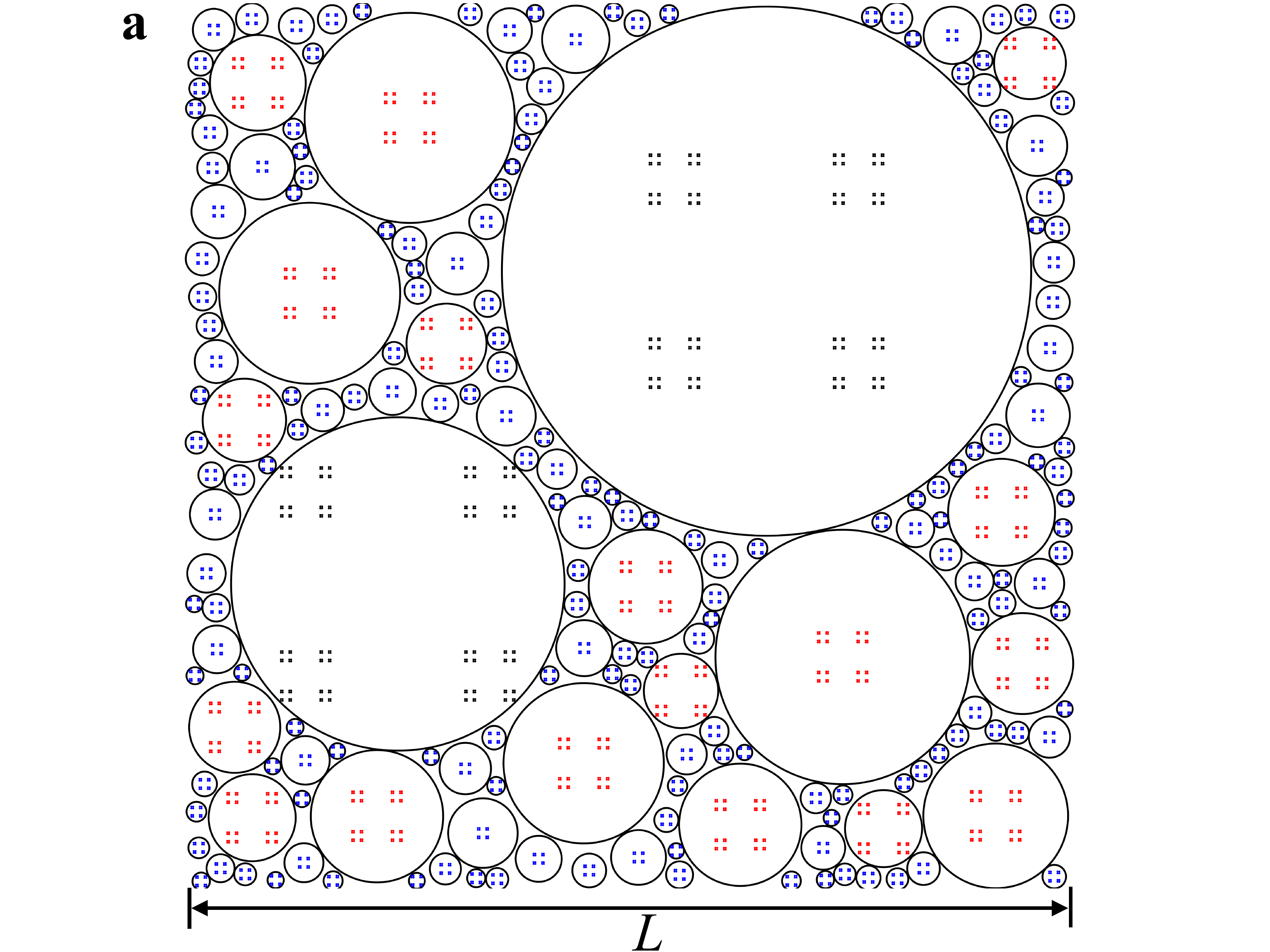}}
\vspace{0.3cm}
\centerline{\includegraphics[width=.9\columnwidth]{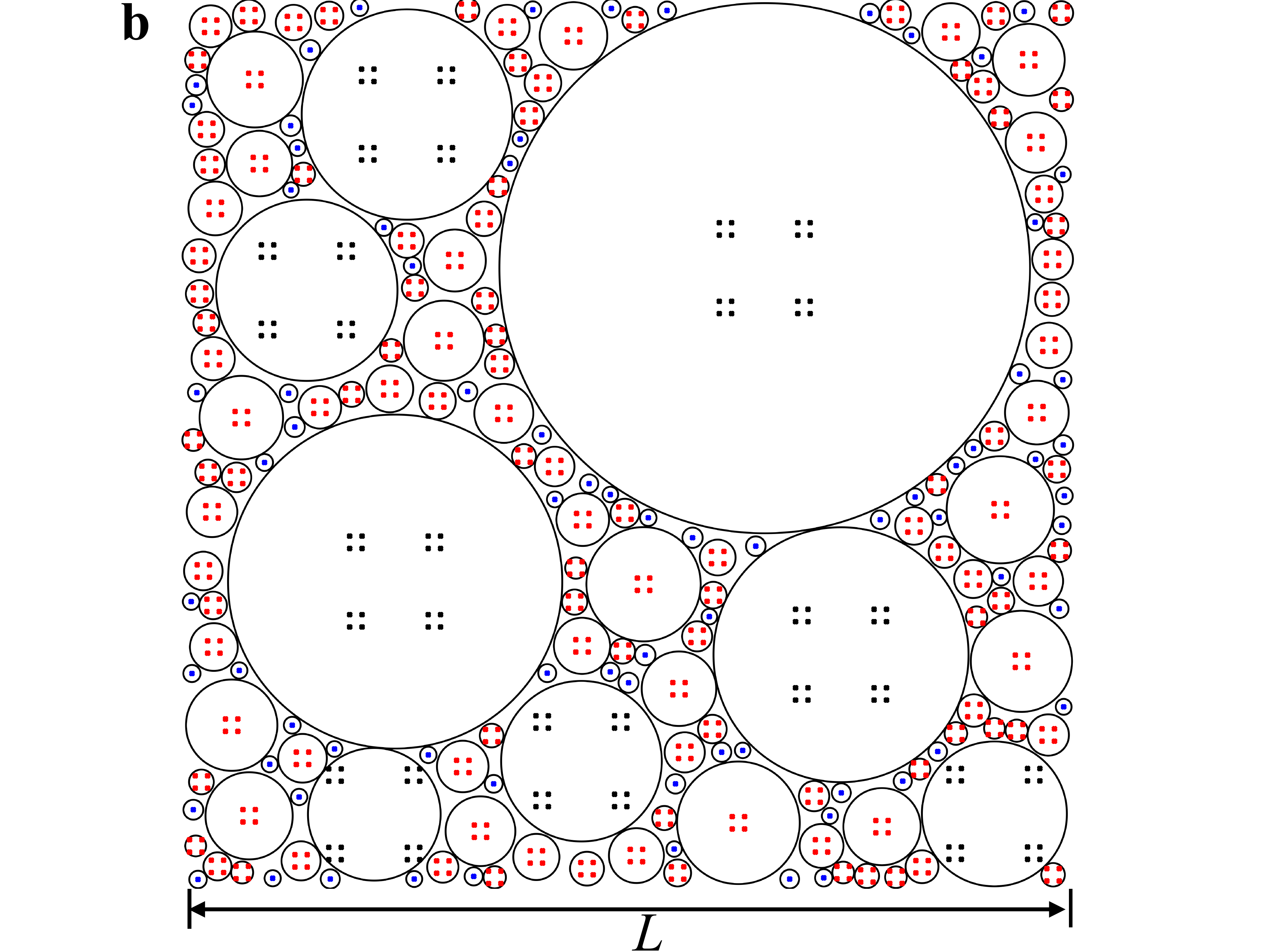}}
\caption{\label{fig:fig11} The model-dependent construction consisting from CMF of fractal dimension $D_{\mathrm{m}}$ inside a power-law distribution of disks (black) with exponent $\gamma = 1.5$, at $f=1$. (a) $D_{\mathrm{m}} = 0.9$ and 2$D_{\mathrm{m}}-\gamma > 0$. (b) $D_{\mathrm{m}} = 0.75$ and 2$D_{\mathrm{m}}-\gamma= 0$.}
\end{figure}

The model-dependent approach is similar to that used for AG and it involves replacing the disks by CMF of dimension $D_{\mathrm{m}}$. Figure~\ref{fig:fig11} shows the model for the two sets of parameters at $f=1$, and with CMF placed inside the first 40 disks, distributed randomly. The radii obey the power-law distribution with the exponent $\gamma = 1.5$. In the construction process, the maximum fractal iteration number of CMF is $m_{\mathrm{max}} = 4$ in Fig.~\ref{fig:fig11}a and $m_{\mathrm{max}} = 3$ in Fig.~\ref{fig:fig11}b. The both structures are depicted in black. Smaller fractal iteration numbers are shown in red ($m=3$ in Fig.~\ref{fig:fig11}a and $m=2$ in Fig.~\ref{fig:fig11}b) and blue ($m=2$ in Fig.~\ref{fig:fig11}a and $m=1$ in Fig.~\ref{fig:fig11}b).

~\begin{figure}[tb!]
\centerline{\includegraphics[width=\columnwidth,trim={0 1cm 2.5cm 2cm}]{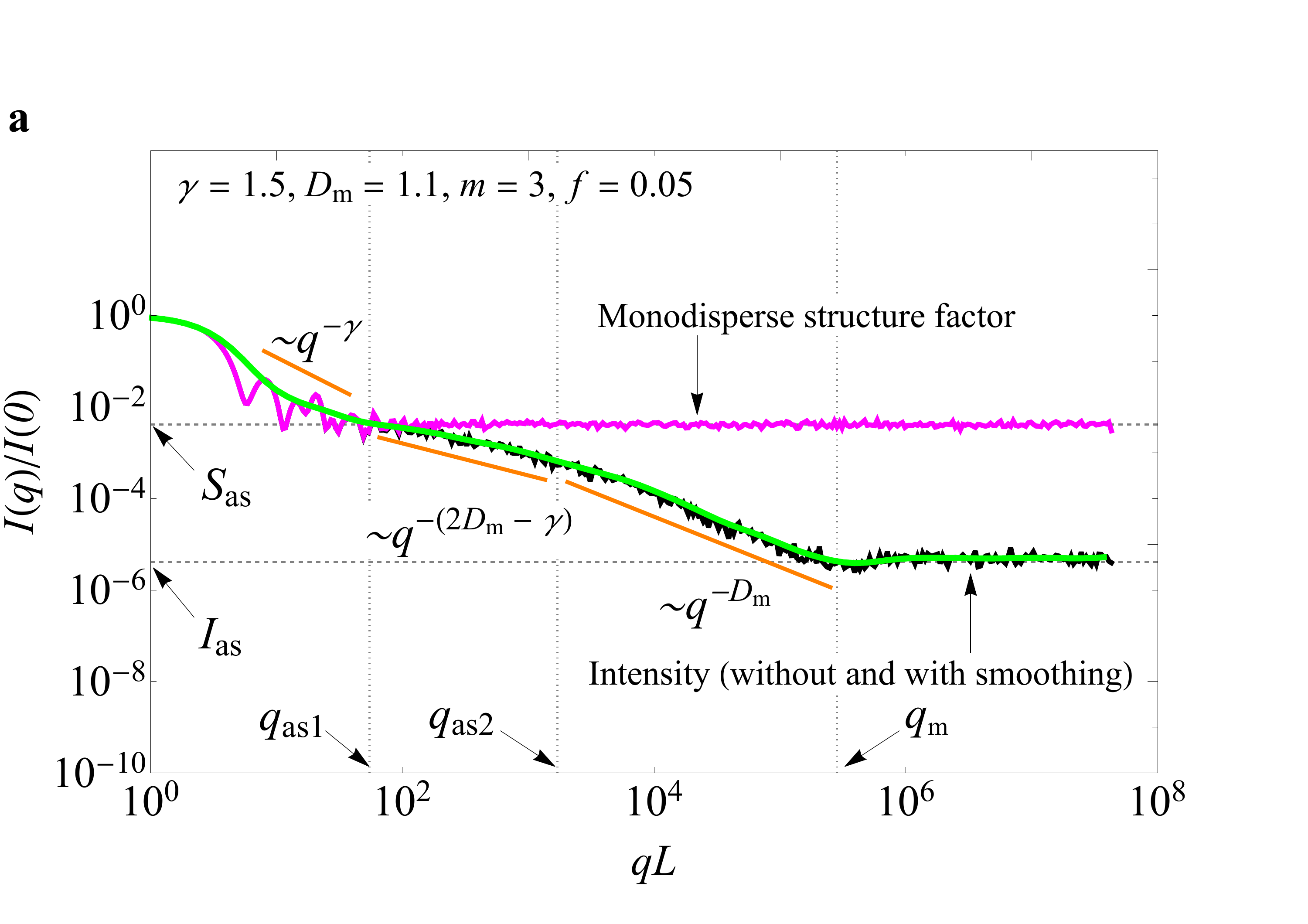}}
\centerline{\includegraphics[width=\columnwidth,trim={0 1cm 2.5cm 2cm}]{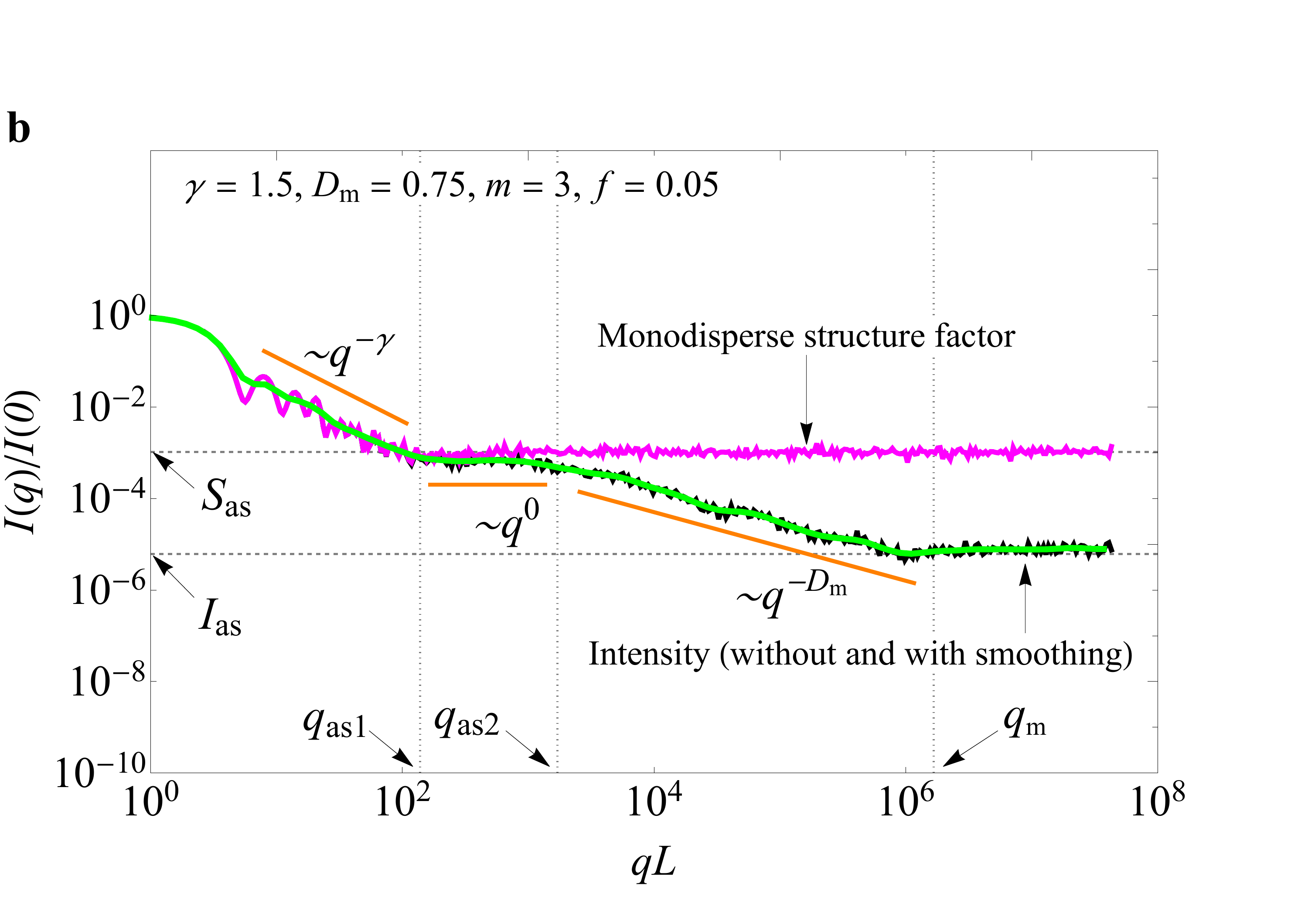}}
\caption{\label{fig:fig12} The scattering curves for the structure shown in Fig.~\ref{fig:fig11}. The notations are the same as in Figs.~\ref{fig:fig5} and~\ref{fig:fig7}. ({a}) ${\gamma} > D_{\mathrm{m}}$ and $2 D_{\mathrm{m}} - {\gamma} > 0$.  ({b}) ${\gamma} > D_{\mathrm{m}}$ and $2D_{\mathrm{m}} - {\gamma} = 0$. The lower dashed horizontal line is $I_{\mathrm{as}} = 1/N_{\mathrm{tot}}$, where $N_{\mathrm{tot}}$ is the total number of scattering points.}
\end{figure}

\subsection{Scattering properties}

Figure \ref{fig:fig12} represents the corresponding scattering intensity and the structure factor at given control parameters.
As in the case of AG consisting of CMF (see Fig.~\ref{fig:fig7}), we recover $\alpha = D_{\mathrm{tot}} = \gamma$ for $\gamma > D_{\mathrm{m}}$ and $2\pi / L \lesssim q \lesssim q_{\mathrm{as1}}$, $\alpha = 2D_{\mathrm{m}} - \gamma$ for $q_{\mathrm{as1}} \lesssim q \lesssim q_{\mathrm{as2}}$, and $\alpha = D_{\mathrm{m}}$ for $q_{\mathrm{as2}} \lesssim q \lesssim q_{\mathrm{m}}$.

\section{Conclusions}
\label{sec:concl}

In Sec.~\ref{sec:entireDim}, we showed that a power-law distribution of fractals forms a fractal-like structure, whose Hausdorff dimension changes provided the exponent $\gamma$ of the distribution is sufficiently big, see Eq.~(\ref{eq:Deff}). Moreover, the condition $\gamma \leqslant d$ is satisfied. Here $d$ is the Euclidean dimension of the embedding space and the upper bound for the exponent $\gamma$ is needed in order to avoid overlapping between fractals.

By using the relations between the fractal dimension and the scattering exponent for mass and surface fractals, we obtained the scattering exponents (\ref{eq:gamM}) and (\ref{eq:gamS}) for power-law polydisperse fractals. The exponent for mass fractals (\ref{eq:gamM}) differs from the exponent found by Martin \cite{martin86} many years ago (see Fig.~\ref{fig:scexp}). In addition, we pointed out that there are restrictions on the resulting exponent, which follow from the restriction imposed on $\gamma$.

In order to verify our predictions, numerical simulations were performed for five models of polydisperse mass fractals: discrete distribution of CMF (Sec.~\ref{sec:models}), two constructions of Apollonian gaskets consisting of the mass fractals (Secs.~\ref{sec:cont1} and \ref{sec:cont2}) and compact packing of power-law polydisperse disks with embedded mass fractals (Secs.~\ref{sec:rpl-disks} and \ref{sec:rpl-CMF}). We obtained that the exponent (\ref{eq:gamM}) is observed just after the Guinier region due to the spatial correlations of mass fractal positions. In the subsequent range of momentum transfer, the spatial correlations decay, and thus the total SAS curve is given by a sum of intensities of separate mass fractals with the exponent \eqref{eq:gamMmar}. Thus, the both our and Martin's exponents are realized but in different ranges of momentum transfer.

We emphasize that the ranges of wave vectors in Figs.~\ref{fig:fig3}, \ref{fig:fig5}, \ref{fig:fig7}, \ref{fig:fig9}, and \ref{fig:fig12} are deliberately chosen to be of 8 orders of magnitude. This is practically not feasible with a single experimental tool, whose range spans about 2 or 3 orders. Then, in practice, any ``window'' of about 2 or 3 orders can be observable, and our purpose is to show all possible behaviour patterns within a narrow region. Note also that the ranges with our and Martin's exponents are located and observed within the first four or five orders of $qL$.

As a prospect, one can study the scattering exponents for dense random packing of power-law polydisperse fractals.

\section{Acknowledgements}

The authors acknowledge support from the JINR--IFIN-HH projects.

\bibliography{pl}

\end{document}